\documentclass[prd,aps,showpacs,superscriptaddress,nofootinbib,amsmath,amssymb]{revtex4}

\usepackage{graphicx}
\usepackage{dcolumn}
\usepackage{bm}%
\usepackage{lipsum}
\begin{document}

\title{Reduction of the Uncertainty in the Atmospheric Neutrino Flux Prediction Below 1 GeV Using Accurately Measured Atmospheric Muon Flux}

\author{M.~Honda}
\affiliation{Institute for Cosmic Ray Research, the University of Tokyo, 5-1-5 Kashiwa-no-ha, Kashiwa, Chiba 277-8582, Japan}
\email[]{mhonda@icrr.u-tokyo.ac.jp}
\homepage[]{http://icrr.u-tokyo.ac.jp/~mhonda}
\author{M.~Sajjad~Athar}
\email[]{sajathar@gmail.com}
\affiliation{Department of Physics, Aligarh Muslim University, Aligarh-202002, India}
  \author{T.~Kajita}
\email[]{kajita@icrr.u-tokyo.ac.jp}
\affiliation{Institute for Cosmic Ray Research, the University of Tokyo, 5-1-5 Kashiwa-no-ha, Kashiwa, Chiba 277-8582, Japan}
\author{K.~Kasahara}
\email[]{kasahara@icrc.u-tokyo.ac.jp}
\affiliation{Shibaura Institute of Technology, 307 Fukasaku, Minuma-ku, Saitama 337-8570, Japan}
\author{S.~Midorikawa}
\email[]{midori@aomori-u.ac.jp}
\affiliation{Faculty of Software and Information Technology, Aomori University,  2-3-1 Kobata, Aomori, Aomori 030-0943, Japan.}
\date{\today}

\begin{abstract}
  We examine the uncertainty of the calculation of the atmospheric neutrino flux
  due to the uncertainty in the hadronic interaction,
  and present a way to reduce it using accurately measured atmospheric muon flux.
  Considering the difference in the hadronic interaction model and  the real one
  as a variation of hadronic interaction, 
  we find a quantitative estimation method for the error of the atmospheric neutrino
  flux calculation from the reconstruction residual of the atmospheric muon flux 
 observed in a precision experiment.
 However, 
 the relation of the calculation error of the  neutrino flux 
 and the reconstruction residual of the  muon flux
 is largely dependent on the atmospheric muon observation site, 
 especially for the low energy neutrinos.
  We study the relation at several observation sites, near Kamioka at sea level,
  same but 2770m a.s.l., Hanle India (4500m a.s.l.), and at Balloon
  altitude ($\sim$ 32km).
  Then, we estimate how stringently the atmospheric muon can reduce 
   the calculation error of the atmospheric neutrino flux.
   We also discuss briefly on the source of error which is considered to be difficult
   to reduce only by the  atmospheric muon data.
 \end{abstract}

\maketitle
\section{Introduction} 
Neutrino oscillation physics has now entered into precision era.
For the atmospheric neutrino, precision  experiments are also planned
at INO~\cite{ino},
South Pole~\cite{pingu},  HyperK~\cite{hyperk}, and DUNE~\cite{dune}, etc.
To address some of the neutrino oscillation parameters,
one requires accurate neutrino flux prediction 
in the $\lesssim$ 1 GeV energy region.

However, it is difficult to calculate the atmospheric neutrino flux
below 1 GeV accurately.
We used to mention that the major source of the uncertainty in
the atmospheric neutrino flux calculation is in those of primary
cosmic ray spectra and hadronic interactions.
Fortunately, with the recent study of primary cosmic ray spectra by
AMS02 and other precision measurements
\cite{ams02p, ams02he, pamela2013, besspolar},
the uncertainty is reasonably reduced to a few \%.
On the other hand the uncertainty of hadronic interaction model
is still large.
Only with the result of high energy experiment, it seems difficult
to reduce the uncertainty to the required level.
On this point we have used the muon flux measured by the precision
experiment to calibrate the hadronic interaction model~\cite{shkkm2006}.
There also are more works having some similarity to this work,
for example, see the
references~\cite{Perkins:1994pm,Perkins:1993zf,Fedynitch2012,Yanez2019}.
Among them, the work in the references~\cite{Fedynitch2012,Yanez2019}
is interesting, since the authors discussed
the uncertainty of hadronic interaction model using observed muon flux
as we do, 
but the target atmospheric neutrino energy range is higher
than that of this paper.

We note that the former studies of hadronic interaction model
using the atmospheric muon for the prediction of atmospheric neutrino
implicitly assume the similarity of the meson production density distribution for
atmospheric neutrino and muon
in the phase space of the hadronic interaction.
For the atmospheric neutrinos with higher energy than a few GeV,
this is true, but, for the atmospheric neutrino below 1 GeV, the situation is
largely different,  due to the energy loss of muon in the atmosphere.
The aim of this paper is to address the effect of the deformation
of density distribution in the phase space of hadronic interaction.

We introduce a mathematical framework for this study in sections~\ref{formulation}
and~\ref{varint},
and try to find an error estimation method of the
uncertainty in the prediction of the atmospheric neutrino flux in section~\ref{variflx}.
In these study, we find the atmospheric muon data is still useful for the
``muon calibration of the hadronic interaction model'' below 1 GeV,
but also find a limitation determined by the muon flux observation site.
We compare the usefulness of the atmospheric muon data observed at
near Kamioka (Tsukuba, sea level and Mt. Norikura, 2770m a.s.l.),
Hanle (India, 4500m a.s.l.)~\cite{hanle}
and at balloon altitude
(near South Pole,  32km a.s.l. by Balloon)~\cite{besspolar}.
We also discuss briefly on the source of error which is considered to be difficult
 to reduce only by the  atmospheric muon data in section~\ref{other}.

\section{\label{formulation}
pseudo-analytic formalism for atmospheric lepton calculation and variation of hadronic interaction model.}

Let us start with a pseudo-analytic expression for the calculation of
atmospheric lepton flux.
It is written as 
\begin{align}
\Phi_L^{obs} ( p_L^{obs}, x^{obs} )  =\sum_{N^{proj}}\sum_{M^{born}} &\int \int 
\Big[ \int  M2L(M^{born},p_M^{born},x^{int},\ L^{obs},p_L^{obs},x^{obs})\notag\\
 &\times H_{int} (N^{proj},p_N^{proj}, M^{born}, p_M^{born})\notag\\
 &\times \sigma^{prod}(N^{proj}, E^{proj})\cdot\rho_{air}(x^{int}) \notag\\
&\times\Phi_{proj}(N^{proj},p_N^{proj},x^{int}) dx^{int} \Big]
\ dp_M^{born} dp_N^{proj} \ ,
\label{eq1}
\end{align}
where $M2L(M,p_M^{born},x^{born}, L^{obs},p_L^{obs},x^{obs})$
is the probability that a meson $M^{born}$ with momentum
$p_M^{born}$ at $x^{born}$ 
decays and result in the lepton $L^{obs}$ with momentum $p_L^{obs}$
at $x_L^{obs}$,
without a hadronic interaction with air nuclei,
$H_{int} (N^{proj},p_N^{proj}, M^{born}, p_M^{born})$ is the probability that
a projectile particle $N^{proj}$ with momentum $p_N^{proj}$ interact with
air nuclei and produce
the $M^{born}$ meson with momentum $p_M^{born}$,
$\sigma^{prod}(N^{proj}, E^{proj})$ is the production cross section
of $N^{proj}$ particle and air nuclei,
$\rho_{air}(x^{int})$ is the nucleus density of the air at $x^{int}$,
and
$\Phi_{proj}(N^{proj},p_N^{proj},x^{int})$ is 
the flux of cosmic ray originated $N^{proj}$-particle 
at $x^{int}$ with momentum $p_N^{proj}$.
Note, we normally use the Monte Carlo simulation to calculate
the atmospheric lepton flux in the actual case.
It is possible to apply this pseudo-analytic  expression to the
real calculation of atmospheric lepton fluxes with a lot of
efforts, but the extension to the
three-dimensional calculation is very difficult.

We use this  pseudo-analytic  expression Eq.~\ref{eq1} to illustrate
the variation study of the hadronic interaction.
With it,  we can close up the hadronic interaction in the atmospheric
lepton flux calculation.
Let us rewrite  Eq.~\ref{eq1} as
\begin{equation}
\Phi_L^{obs} ( p_L^{obs}, x^{obs} )  =\sum_{N^{proj}}\sum_{M^{born}} \int \int \ 
D(N^{proj},p_N^{proj}, M^{born}, p_M^{born},\ L^{obs},p_L^{obs},x^{obs}) 
\ dp_{M}^{born} dp_{N}^{proj} \ ,
\label{eq2}
\end{equation}
and
\begin{align}
D(N^{proj},p_N^{proj}, M^{born}, p_M^{born},\ L^{obs},p_L^{obs},x^{obs}) 
=& \int M2L(M^{born},p_M^{born},x^{int},\ L^{obs},p_L^{obs},x^{obs})\notag\\
  &\times H_{int} (N^{proj},p_N^{proj}, M^{born}, p_M^{born})\notag\\
 &\times \sigma^{prod}(N^{proj}, E^{proj})\cdot\rho_{air}(x^{int}) \notag\\
  &\times\Phi_{proj}(N^{proj},p_N^{proj},x^{int})\ \ dx^{int} \ \ ,
\label{eq3}
\end{align}
The D-function in Eq.~\ref{eq2} is the density distribution for atmospheric lepton
in the phase space of the hadronic interaction.
We call it as the ``integral kernel'' of atmospheric lepton flux.
Classifying the projectile particle into three categories;
proton, neutron, and all mesons,
we consider the integral kernel for all combination of those projectile
and the secondary mesons whose decay branching ratio to leptons
or semi-leptonic decay is larger than 1~\%
($\pi^\pm, K^\pm$, and $K^0_L$).
Note, a  nucleus projectile hadronic interaction is normally represented
by the superposition of single nucleon interactions.
Adding to these nucleon projectiles, the meson created in the hadronic
interaction with air nuclei also can be the projectile in the next interaction.
However, 
the meson projectiles (mainly $\pi^\pm$) are not important yet in the
energy region we are working due to their short life time,
then we summarize them in a category.

As we mentioned above, the atmospheric lepton flux is normally
calculated with the Monte Carlo simulation, we calculate the integral kernel
with the Monte Carlo simulation.
The Monte Carlo simulation we use here is the same one 
used in our calculation of atmospheric neutrino and 
muon fluxes~\cite{hkkm2011, hakkm2015}.
We tag all the particles appeared in the simulation,
and record the projectile particle and the secondary meson momenta,
when the meson create the target lepton without hadronic interaction.
Then we study the 
($p_M^{born}, p_N^{proj}$) point distribution in the hadronic interaction
phase space.

\begin{figure}[ht]
  \includegraphics[height=6.8cm]{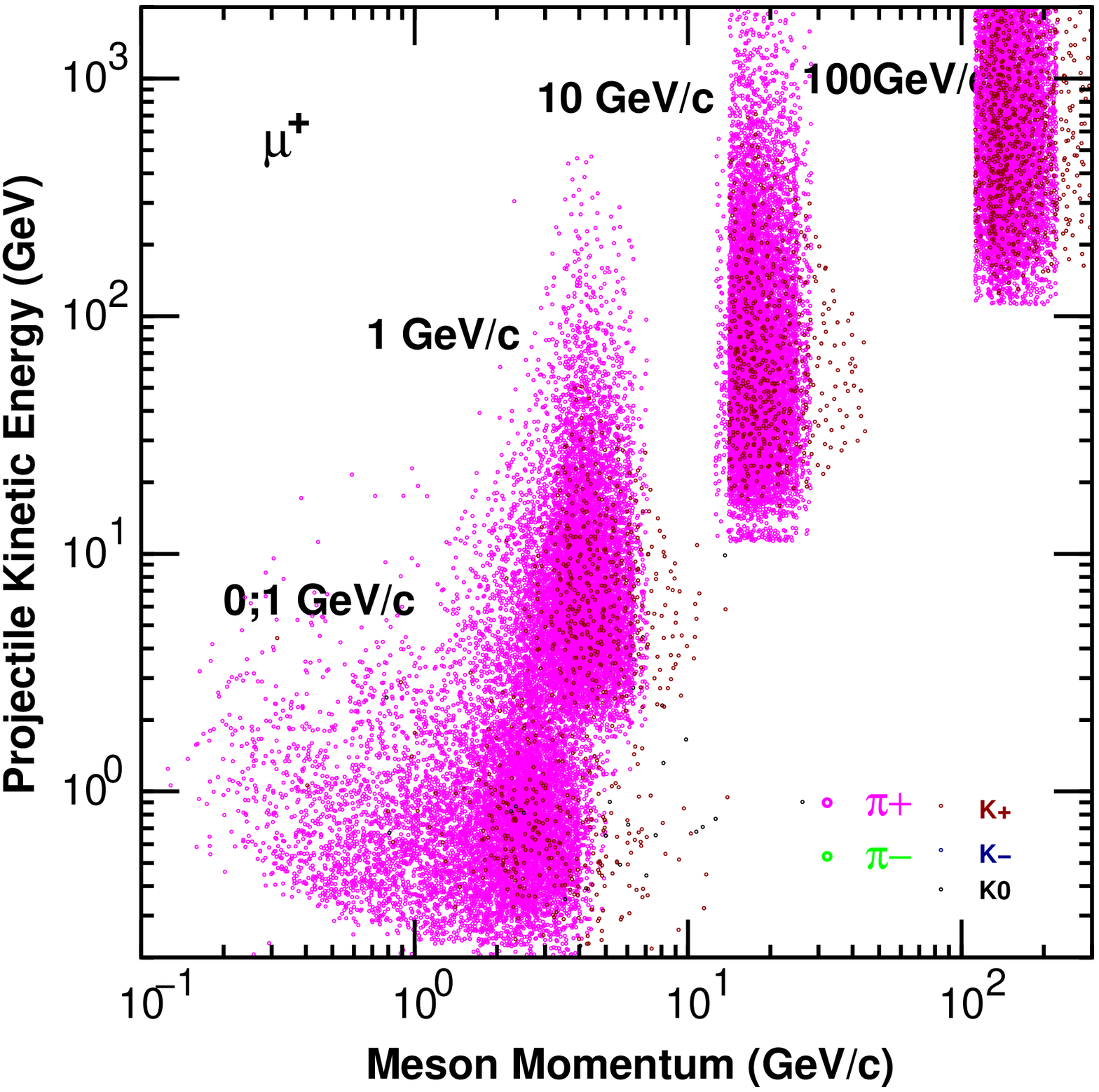}
  \hspace{5mm}%
 \includegraphics[height=6.8cm]{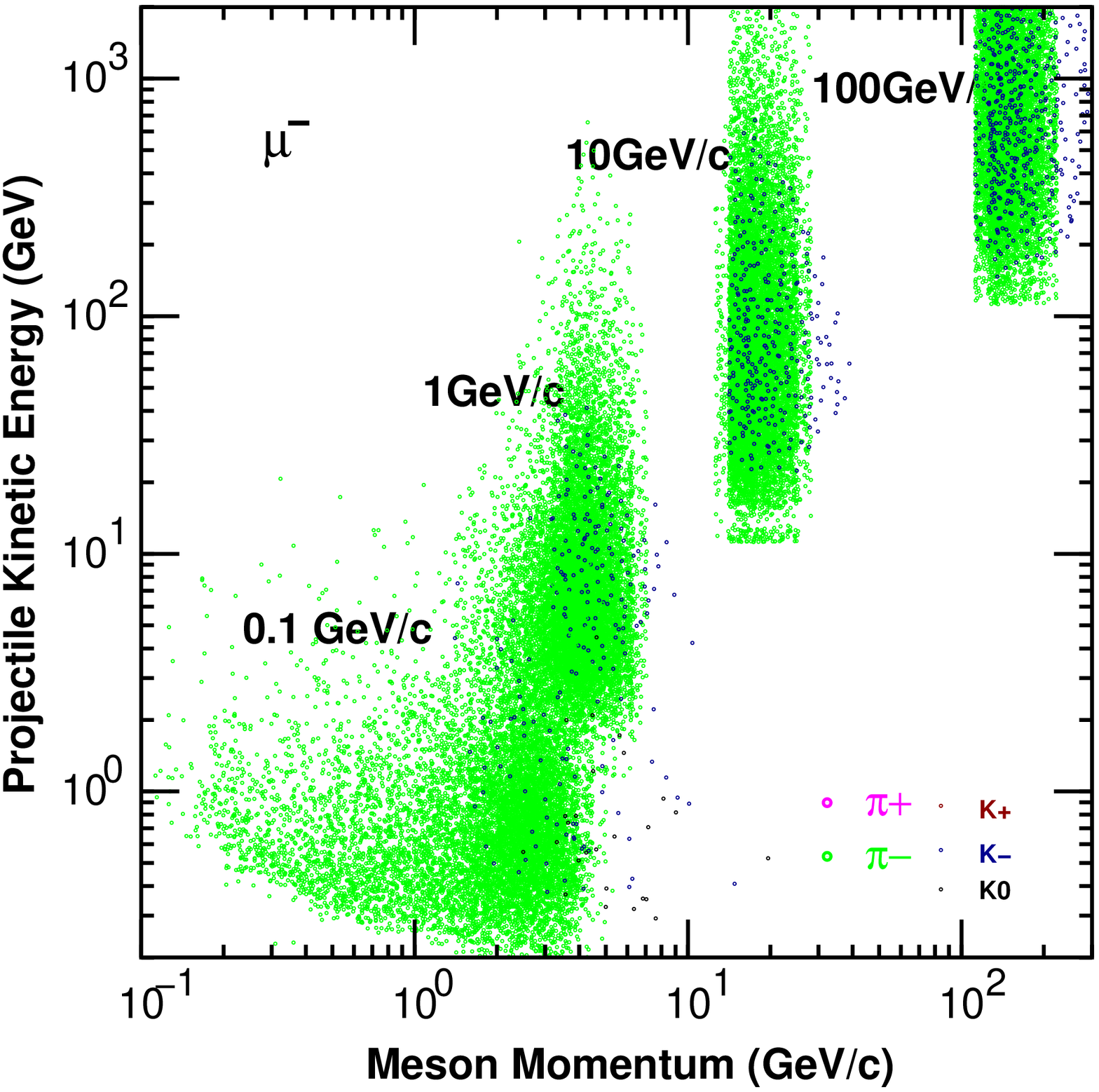}
  \caption{\label{mu-scatter}
    The scatter plot of the projectile and meson momenta, which
    create 0.1, 1.0,  10 and 100 GeV/c downward moving atmospheric muon
    at Kamioka (sea level),
    in the hadronic interaction phase space.
    We plot those of $\mu^+$ in the left  panel, and  $\mu^-$ in the right panel.
    Note, we use the kinetic energy for the projectile particle in the figure
    to magnify the region below 1 GeV/c.
  }
\end{figure}

\begin{figure}
\includegraphics[height=6.8cm]{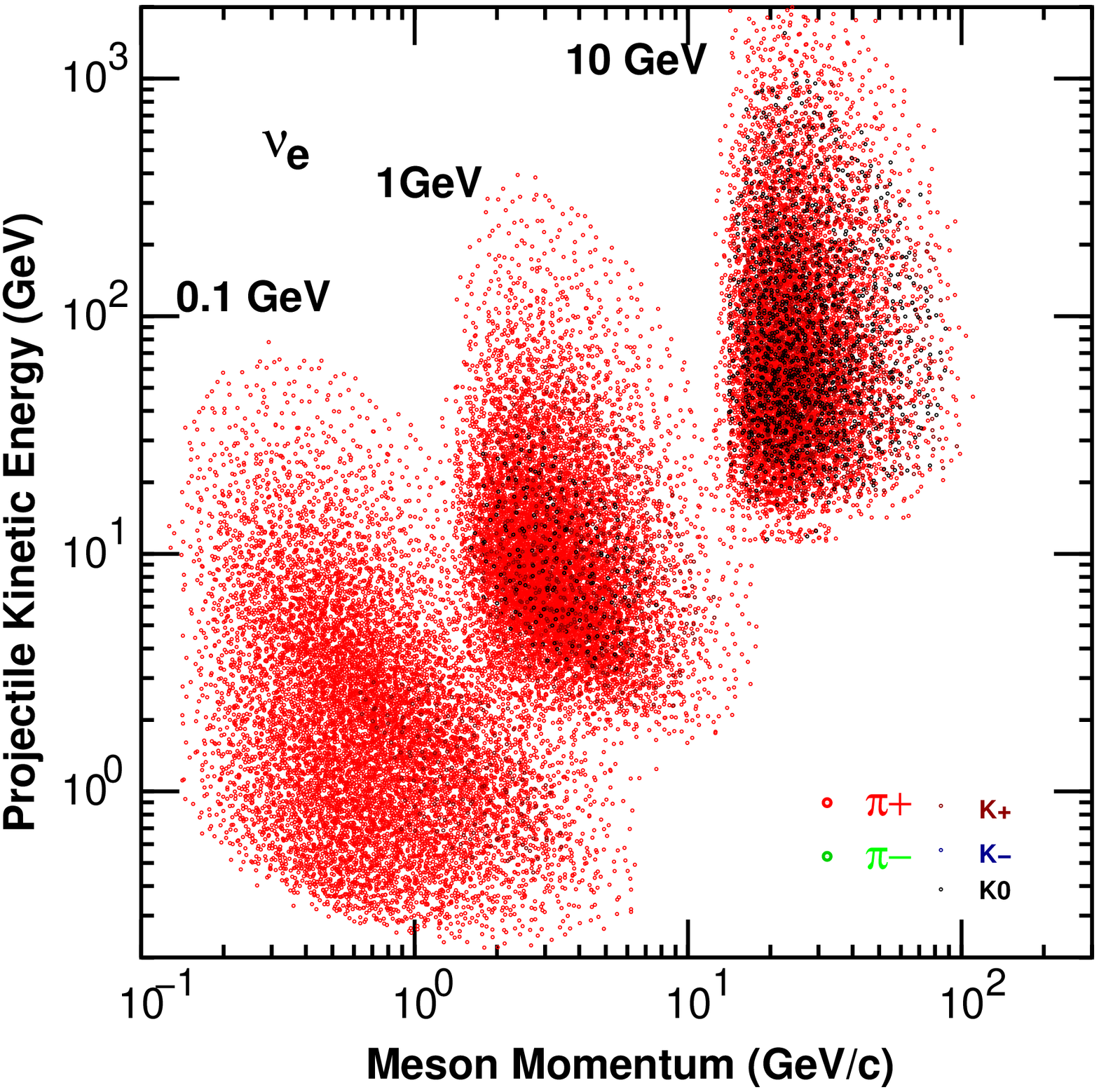}%
\hspace{5mm}%
\includegraphics[height=6.8cm]{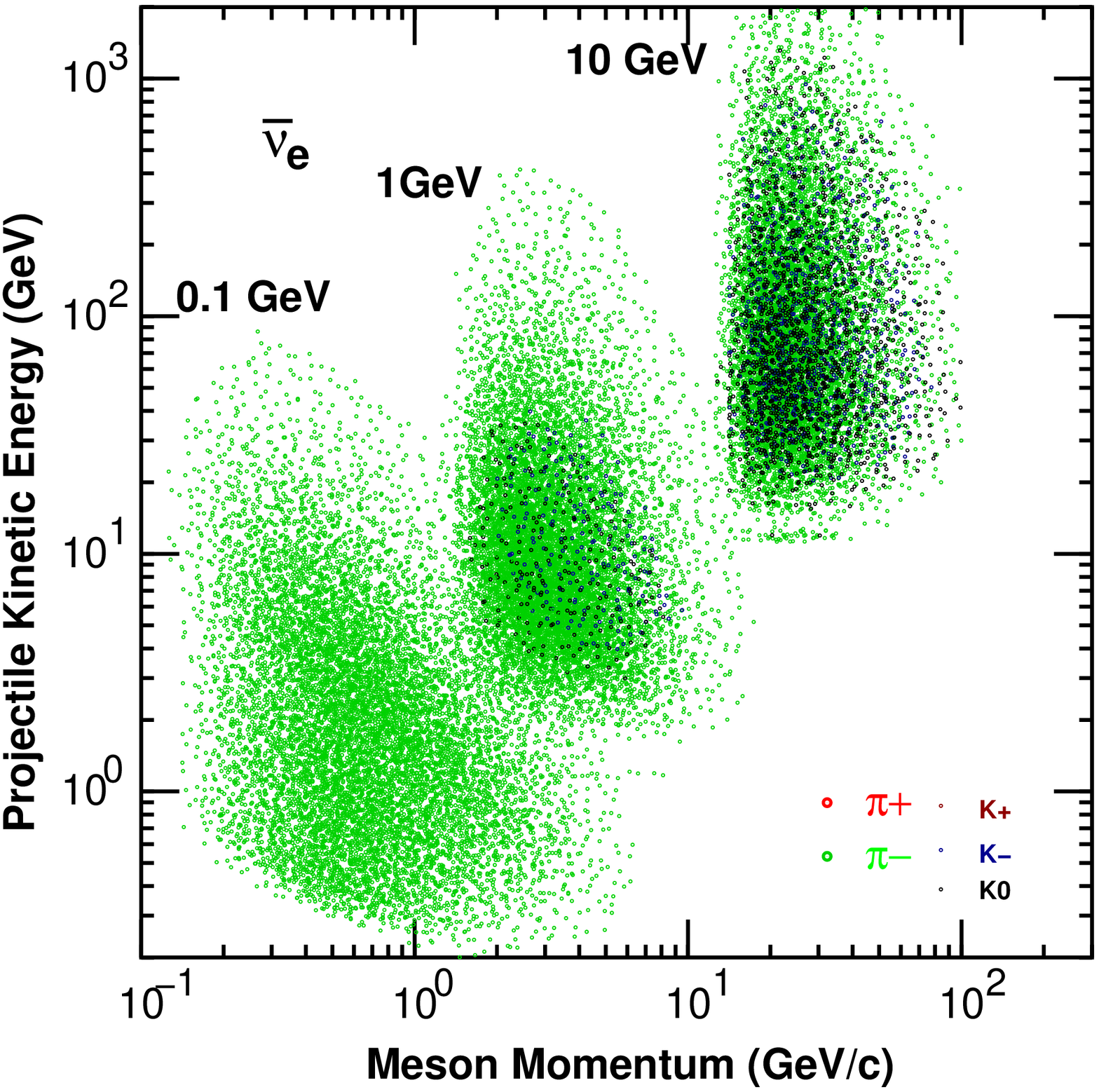}%
\\
\includegraphics[height=6.8cm]{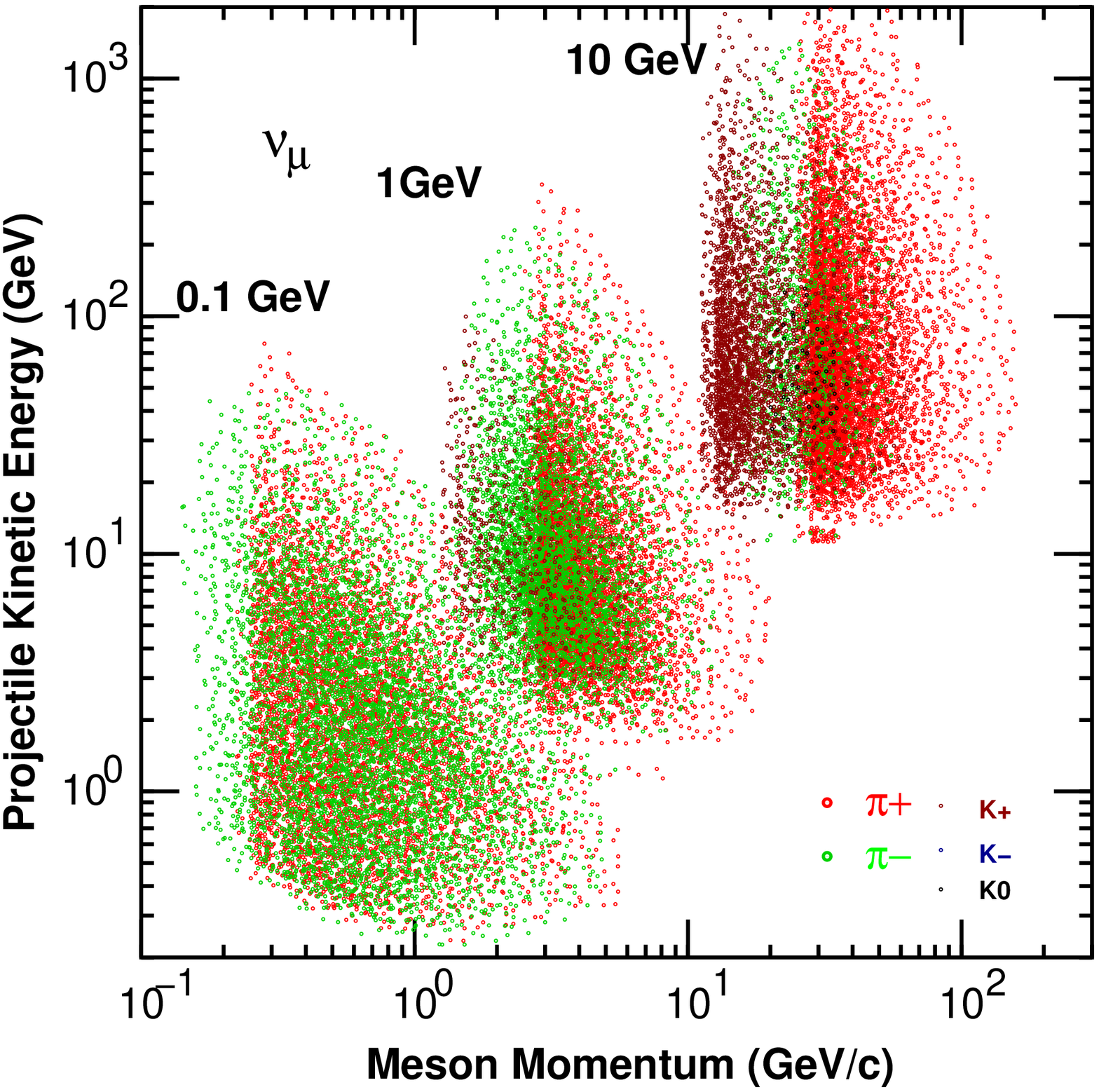}%
\hspace{5mm}%
\includegraphics[height=6.8cm]{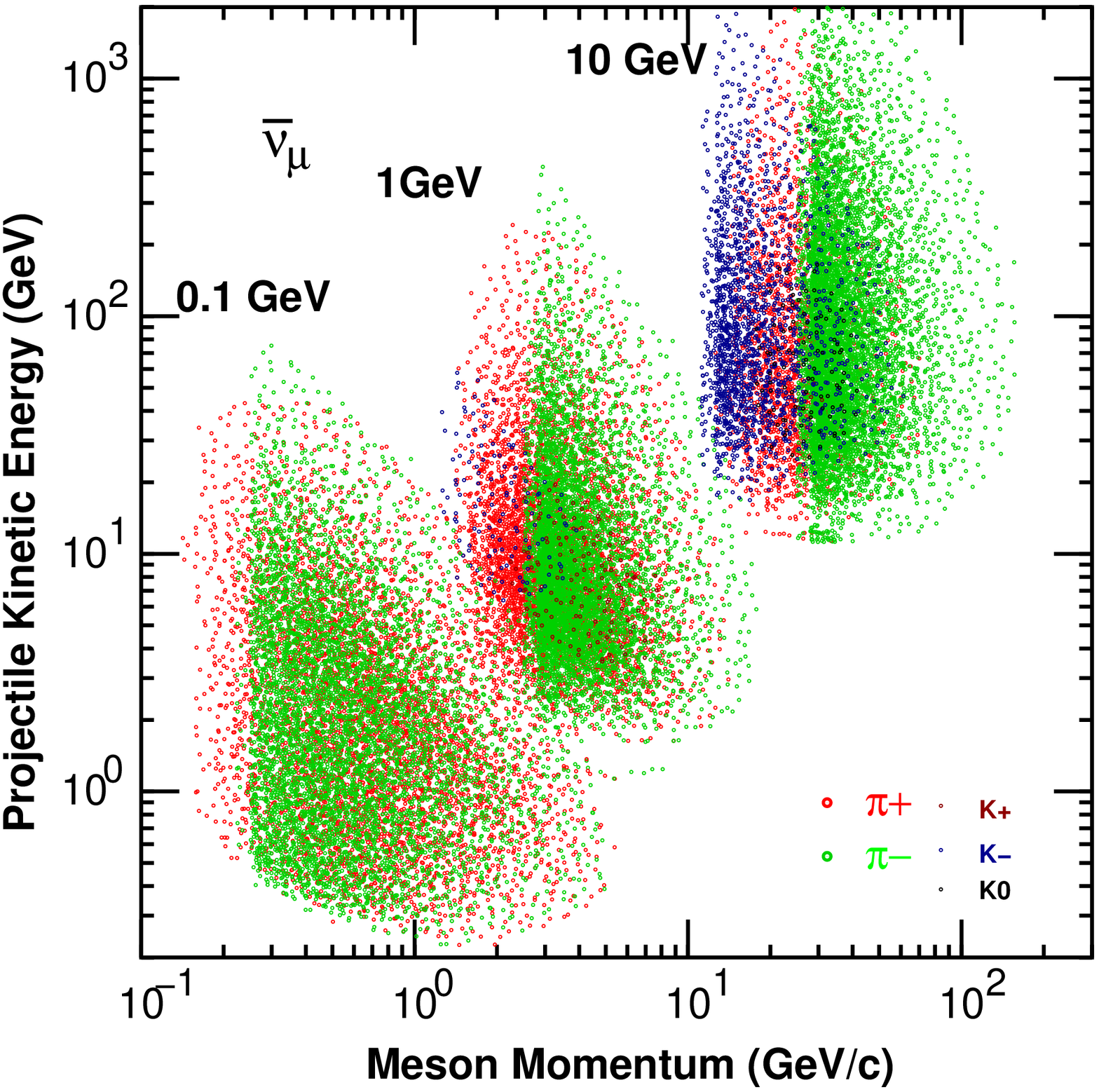}%
\caption{\label{nu-scatter}
  The scatter plot of the  projectile and meson momenta, which
  create the 0.1, 1.0, and 10 GeV vertically downward moving neutrino
  at Kamioka (sea level),
  in the hadronic interaction phase space.
  We plot those of $\nu_e$ in the top left  panel,   $\bar\nu_e$ in the top right panel,
  $\nu_\mu$ in the bottom left  panel and  $\bar\nu_\mu$ in the bottom right panel.
  Note, we use the kinetic energy for the projectile particle in the figure
  to magnify the region below 1 GeV/c.
}
\end{figure}

The full three-dimensional Monte Carlo simulation for atmospheric neutrino
need a long computation time, since it is an Earth size simulation
for upward moving neutrino.
However, if we limit the calculation to the  downward going neutrino only, 
it becomes far less time consuming simulation.
We consider here only the downward moving atmospheric neutrino
 as well as the muon.
 As the examples, we show the integral kernel as the scatter plot
 in Fig.~\ref{mu-scatter}  
 for the 0.1, 1.0, 10  and 100 GeV/c vertically downward moving muon at Kamioka
 (sea level),
and in Fig.~\ref{nu-scatter} 
for 0.1, 1.0 and 10  GeV vertically downward moving neutrino at Kamioka.
Note, we use the kinetic energy for the projectile particle in the figure
to magnify the region below 1 GeV/c.
Also, we plot all the ($p_M^{born}, p_N^{proj}$)  points by
different projectiles  ($p,n,mesons$) in the same figure.

We find the integral kernels for atmospheric neutrino and muon 
moves almost parallel with their momentum or energy above
1 GeV/c for atmospheric muon and above 1 GeV for atmospheric
neutrino.
However, the integral kernels for atmospheric muon show a
large deformation at 0.1 GeV/c, and the central momentum of the
parent meson is very close to the atmospheric muon at 1 GeV.
On the other hand, the integral kernel of the atmospheric neutrino
at 0.1 GeV shows a little deformation, but keeps the similarity to
that of higher energies.

The integral kernel is sensitive not only to the lepton momentum,
but also to the direction of the lepton motion, and to the observation site,
especially for the atmospheric muon flux.
We calculate the integral kernel of atmospheric neutrino flux
for vertical downward and horizontal directions at Kamioka, 
and that of muon flux
for vertical downward and horizontal directions at several observation sites
including Kamioka.

\section{\label{varint}
  variation of the interaction model with random numbers.}

If we assume  the projectile flux
$\Phi_{proj}(N^{proj},p_N^{proj},x^{int})$
is not largely affected by the variation of the hadronic interactions,
we can study the effect of the variation of the hadronic interactions
on the lepton flux using the pseudo-analytic formalism.
The lepton flux calculated by the varied interaction model
may be written as;
\begin{align}
\tilde \Phi_L^{obs} ( p_L^{obs}, x^{obs} )  =&\sum_{N^{proj}}\sum_{M^{born}} \int \int 
\Big[ \int  M2L(M^{born},p_M^{born},x^{int},\ L^{obs},p_L^{obs},x^{obs})\notag\\
  &\hspace{2cm} \times H_{int} (N^{proj},p_N^{proj}, x^{int}, M^{born}, p_M^{born})
  \cdot \left(1 + \Delta_{int} (N^{proj}, M^{born}, p_N^{proj},p_M^{born}) \right)\notag\\
 &\hspace{2cm}\times \sigma^{prod}(N^{proj}, E^{proj})\cdot\rho_{air}(x^{int}) \notag\\
  &\hspace{2cm}\times\Phi_{proj}(N^{proj},p_N^{proj},x^{int}) dx^{int} \Big] \ dp_M^{born} dp_N^{proj} \\
  =&\sum_{N^{proj}}\sum_{M^{born}} \int \int \ 
D(N^{proj},p_N^{proj}, M^{born}, p_M^{born},\ L^{obs},p_L^{obs}) \notag\\
&\hspace{2cm}\times \left(1 + \Delta_{int} (N^{proj}, M^{born}, p_N^{proj},p_M^{born}) \right)
\ dp_M^{born} dp_N^{proj} \ .
\label{general-df}
\end{align}
We can use Eq.~\ref{general-df} 
to construct a variation of interaction model and the variation
of atmospheric lepton fluxes and study it.

The variation of the hadronic interaction model with random numbers
can be constructed with the help of the B-spline functions.
We use the 3rd order B-spline function with constant knot separation, and is
represented as
\begin{equation}
  B_\Delta^{i}(x) =  b\left({{x-i \cdot \Delta}\over\Delta} -x_0\right)   
\label{bspl0}
\end{equation}
where
\begin{align}
b(t) =\left\{
\begin{array}{ll}
  \ \ {1 \over 6}(3|t|^3 - 6 |t|^2 +4)
  & (|t| \leq 1)\\
  \\
  -{1 \over 6}(|t| - 2)^3
  &(1\leq |t| \leq 2)\\
  \\
 \ \ \ 0 & ( |t| \geq 2)\\
\end{array}
\right.\ \ \,
\label{bspl1}
\end{align}
where $\Delta$ is the knot separation, and $x_0$ is the origin, 
normally taken as  $x_0=0$.
The linear combination of the 3rd order B-spline function (Eq.~\ref{bspl0} and Eq.~\ref{bspl1})
is continuous up to the 2nd order derivative, and  is often used to connect the
discrete data or to fit them.

Using the B-spline function, we construct the variation function as
\begin{equation}
  \Delta_{int} (N^{proj}, M^{born}, p_N^{proj},p_M^{born}) = \delta \cdot
  \sum_i \sum_j R_N^{ij}
  \cdot B_{\Delta_{proj}}^{i} ( \log_{10}(p_N^{proj}))
  \cdot B_{\Delta_{meson}}^{j} ( \log_{10}(p_M^{born})) \  ,
  \label{bspl-dint}
\end{equation}
Then we can write the variation of lepton flux 
calculated with this variation of interaction model as,
\begin{align}
  \tilde \Phi_L^{obs} ( p_L^{obs}, x^{obs} )
    =&\sum_{N^{proj}}\sum_{M^{born}} \int \int \ 
D(N^{proj},p_N^{proj}, M^{born}, p_M^{born},\ L^{obs},p_L^{obs}) \notag\\
&\hspace{0.5mm}\times \left(1 +
\delta\sum_i \sum_j R_N^{ij}\cdot
  B_\Delta^{i} ( \log_{10}(p_N^{proj}))
  \cdot B_\Delta^{j} ( \log_{10}(p_M^{born}))
  \right) dp_M^{born} dp_N^{proj} \ ,
  \label{bspl-fd}
\end{align}
and the variation of the lepton flux as
\begin{align}
  \Delta \Phi_L^{obs} ( p_L^{obs}, x^{obs} ) \equiv& 
  \tilde \Phi_L^{obs} ( p_L^{obs}, x^{obs} ) -
  \Phi_L^{obs} ( p_L^{obs}, x^{obs} ) \notag\\ 
    =&\delta \sum_{N^{proj}}\sum_{M^{born}} \int \int \ 
D(N^{proj},p_N^{proj}, M^{born}, p_M^{born},\ L^{obs},p_L^{obs}) \notag\\
&\hspace{5mm}\times 
\sum_i \sum_j R_N^{ij}\cdot
   B_\Delta^{i} ( \log_{10}(p_N^{proj}))
  \cdot B_\Delta^{j} ( \log_{10}(p_M^{born}))
dp_M^{born} dp_N^{proj} \  .
\label{bspl-df}
\end{align}
Here, we assume $\{  R_N^{ij} \}$  as the set of  normal random numbers
with the average value = 0 and the standard deviation = 1, which
is one of the standard random number in the computer science.
We take $\Delta_{proj} = \Delta_{meson} = \Delta (=0.5) $ in Eq.~\ref{bspl-fd}.
This means we consider the variation of interaction model in the momentum scale
$\Delta \log_{10}(p)\gtrsim 0.5$ both for the projectile and secondary meson momenta.

When the random number set $\{  R_N^{ij} \}$ is given,
the variation of the integral kernel density at a grid point $\{ij\}$ is written as
\begin{align}
  \Delta D_{ij}
  = D_{ij}
  \times \delta
\sum_k \sum_l 
 R_N^{kl}\cdot
   B_\Delta^{k} ( (k-i)\cdot \Delta)
  \cdot B_\Delta^{l} ((l-j)\cdot\Delta) \ ,
\label{Ddens}
\end{align}
where we have simplified the kernel density at the grid point 
$D(N^{proj},(p_N^{proj})_i, M^{born}, (p_M^{born})_j,\ L^{obs},p_L^{obs}) $ as
$D_{ij}$.
Since $\{R_N^{kl}\}$ are the set of independent normal random numbers with
average value 0, and standard deviation 1,
the variance  or the square of the standard deviation of $  \Delta D_{ij} $ is
calculated as
\begin{align}
  \sigma_{D_{ij}}^2=D^2_{ij}
  \times \delta^2
\sum_k \sum_l 
   \left[B_\Delta^{k} ( (k-i)\cdot \Delta)  \cdot B_\Delta^{l} ((l-j)\cdot\Delta)\right]^2 \  .
\label{Ddens2}
\end{align}
With the definition of B-spline function (Eq.~\ref{bspl0}  and Eq.~\ref{bspl1}),
the equation is easily evaluated as,
\begin{equation}
  \sigma^{\ }_{D_{ij}} = 0.5\cdot \delta \cdot  D_{ij} \ .
  \label{Ddens3}
\end{equation}
Note, 
we apply an independent set of random numbers to the integral kernel
calculated for each combination of all the projectile and all the
secondary meson.

\begin{figure}[ht]
\includegraphics[height=5.5cm]{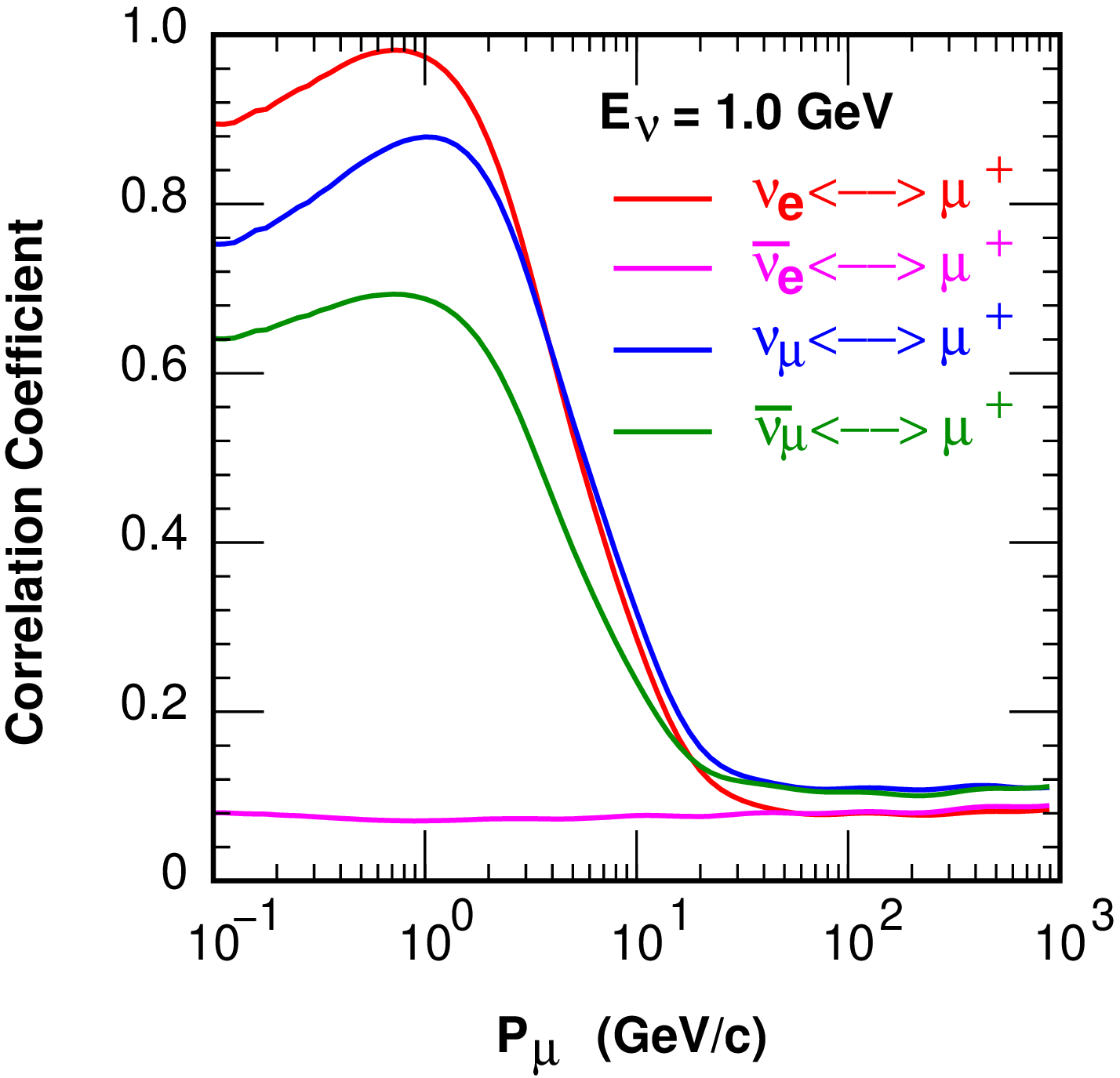}\hspace{5mm}%
\includegraphics[height=5.5cm]{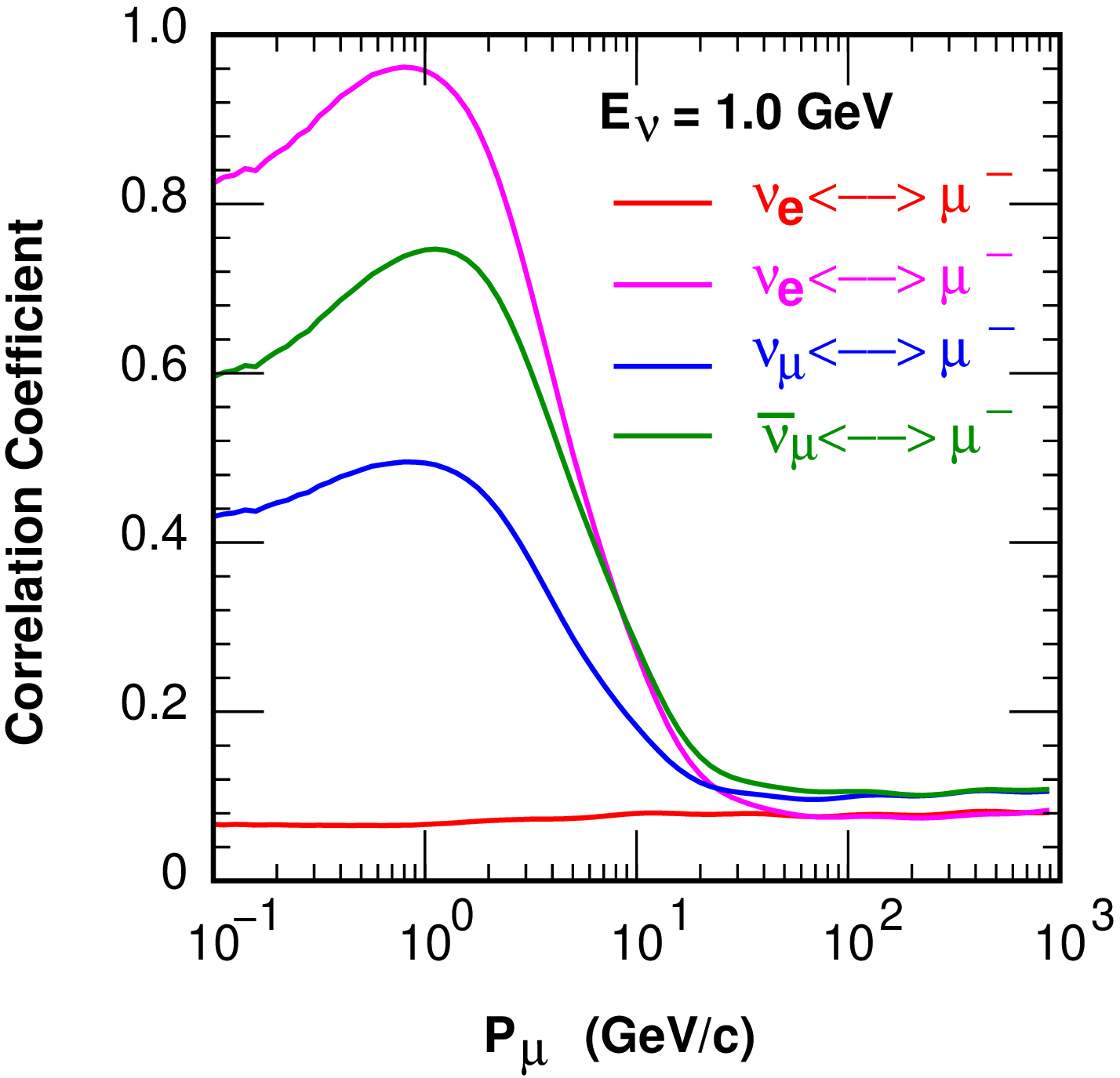}%
\caption{\label{corr1}
  Correlation coefficient for each combination of neutrinos
  ($\nu_\mu, \bar\nu_\mu,\nu_e,\bar\nu_e$) and muons ($\mu^+,\mu^-$)
  tat neutrino energy of 1 GeV.
  We used the integral kernels 
  of atmospheric neutrinos and muons
  both for vertically downward moving ones at Kamioka.
}
\end{figure}

As an application of the variation of the interaction model with random number,
we calculate the correlation coefficient of atmospheric neutrino and atmospheric muon fluxes as,
\begin{equation}
  \gamma( p_\nu^{obs}, x_\nu^{obs}; p_\mu^{obs}, x_\mu^{obs}) =
      {
          {
            \sum \left(\Delta \Phi_\nu(p_\nu^{obs}, x_\nu^{obs})
            \Delta\Phi_\mu(p_\mu^{obs}, x_\mu^{obs})\right)
          }
    \over
    \sqrt{
      \sum\left(\Delta \Phi_\nu^k(p_\nu^{obs}, x_\nu^{obs})\right)^2
      \sum\left(\Delta\Phi_\mu^k(p_\mu^{obs}, x_\mu^{obs})\right)^2
    }
        }
  \label{corrcoef}
\end{equation}
and study the correlation coefficient between muon and neuron fluxes
at each combination of muon momentum and neutrino energy.
As an example, we show the correlation coefficient of neutrino flux
at 1 GeV and the muon fluxes as the function of muon momentum
in fig.~\ref{corr1} for all combination of
($\nu_\mu, \bar\nu_\mu,\nu_e,\bar\nu_e$) and ($\mu^+,\mu^-$).
Here we used the integral kernel of the vertically downward moving fluxes 
of atmospheric neutrinos and muons at Kamioka.
Note, 
$\pi^-$  creates
almost all of $\bar\nu_e $ and $\mu^-$, and $\pi^+$ creates
almost all of  $\nu_e$ and $\mu^+$ in their decay cascade;
\begin{align}
  \pi^{+(-)} \rightarrow &\mu^{+(-)} + \nu_\mu (\bar\nu_\mu) \notag\\
  &\downarrow\notag\\
  &e^{+(-)} + \bar\nu_\mu (\nu_\mu) + \nu_e(\bar\nu_e)
\label{decay}
\end{align}
at low energies.
The correlation coefficient of  electron neutrino 
and  muon fluxes created by different types of pion are small and 
no meaningful structure is seen as the function of muon momentum
in Fig.~\ref{corr1}.
On  the other hand, in the case of muon neutrinos,
$\pi^+$ creates
$\nu_\mu$, $\bar \nu_\mu$, and $\mu^+$, and
$\pi^-$ creates
$\nu_\mu$, $\bar \nu_\mu$, and $\mu^-$.
Therefore both signed muon have correlation to both
type of muon neutrinos.

\begin{figure}[ht]
\includegraphics[height=5.5cm]{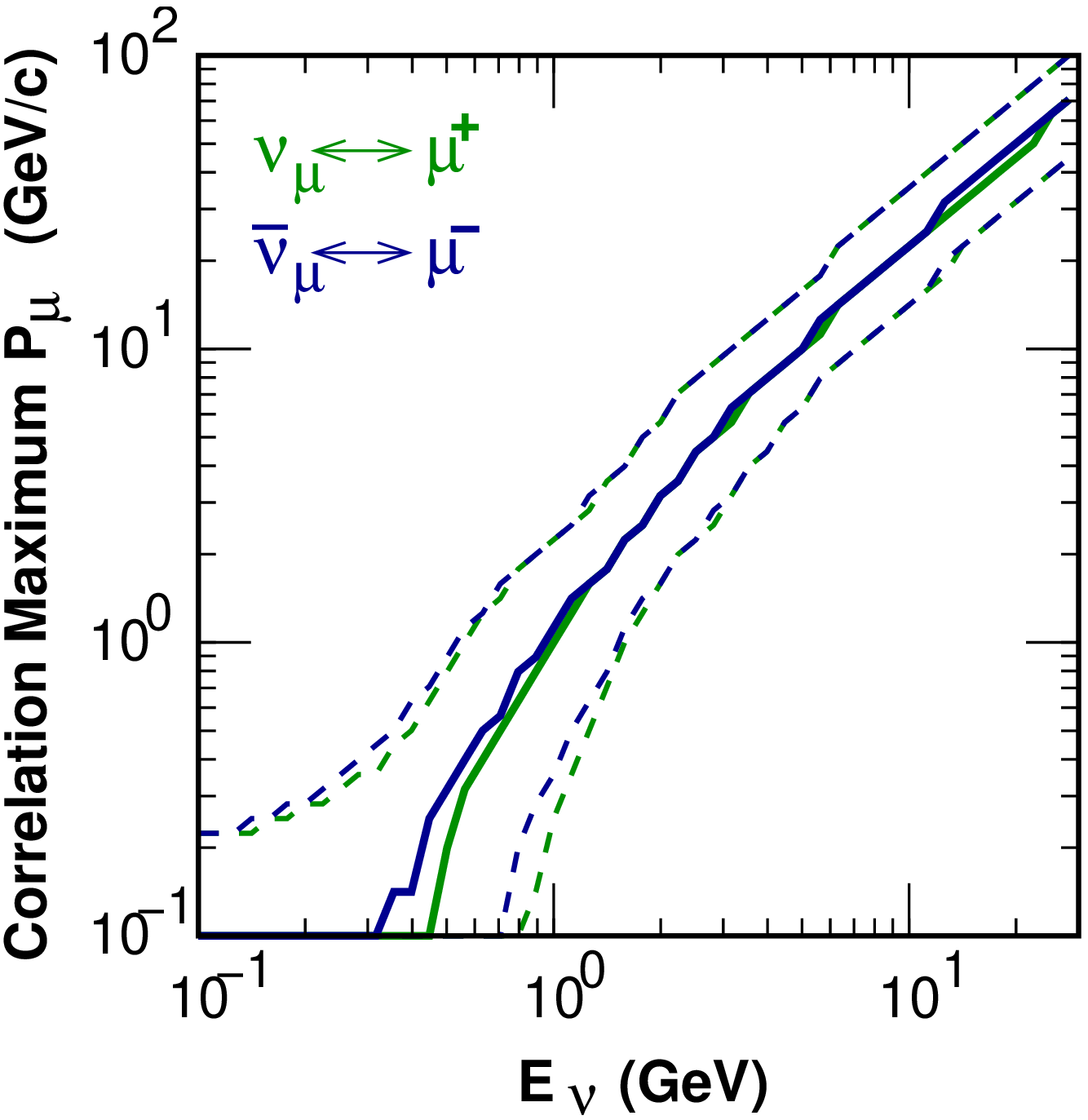}\hspace{5mm}%
\includegraphics[height=5.5cm]{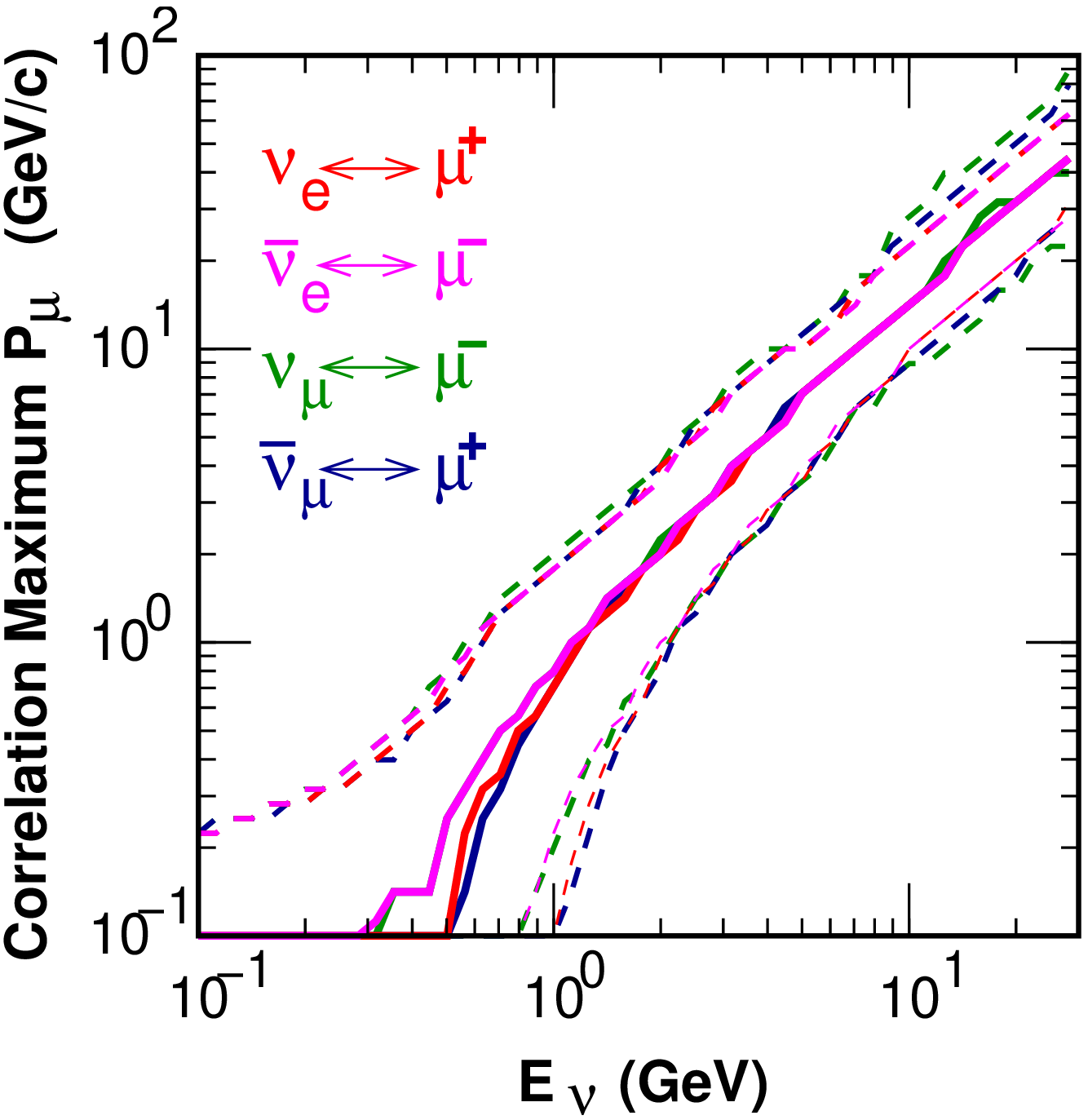}%
\caption{\label{EPcorr}
  The muon momenta at which  the muon show the maximum correlation and
the 90\% of that  to the neutrino at fixed energy (horizontal axis).
 The left panel is for the neutrinos directly produced at the pion decay, and in the
  right panel for the neutrinos produced at the decay of muon
  which produced by the pion decay.
  We used the integral kernels 
  of vertically downward moving atmospheric neutrinos and muons
  at Tsukuba or Kamioka (sea level) 
}
\end{figure}

In Fig.\ref{EPcorr}, 
we show the muon momentum which gives the
maximum correlation coefficient and 90\% of it
as the function of neutrino energy
for the direct decay product of $\pi^\pm$
(left panel)
and decay product of $\mu^\pm$ (right panel)
separately, to see the difference due to the
decay kinematics.
Note, we do not show the plot for $\nu_\mu \leftrightarrow \mu^-$
and  $\bar\nu_\mu \leftrightarrow \mu^+$,
since there are no meaningful correlation between them
(Fig.~\ref{corr1}).
In  both the panels, we find the lines for maximum correlation
and 90~\% of it are very close among different type of the neutrinos,
but with the same kinematics.

\section{\label{variflx}
  Variation of  atmospheric neutrino  and muon fluxes}

In this section, we generate a huge number (3,000,000) of the normal random number sets,
and study the variation of atmospheric neutrino flux when  the variation
of atmospheric muon flux is limited.
To cover a large variation of the interaction model at the beginning,
we take $\delta=1$ in Eq.\ref{bspl-dint} in this section.
After fixing the kind of target neutrino and it's energy,
we calculate the correlation coefficient of both signed atmospheric muon flux
to the target neutrino as the function of the muon momentum.
For each signed muon flux,
when it has a meaningful correlation maximum,
we put the constraint on the  flux variation to satisfies the condition;
\begin{equation}
 \left|{ {\Delta \phi_\mu} \over \phi_\mu}\right|  < \varepsilon \ ,
\label{mulimit}
\end{equation}
in the momentum range where the correlation coefficient is larger than
the 90~\% of the maximum.
Therefore, 
when the target neutrino is electron neutrino ($\nu_e$ and $\bar\nu_e$),
the flux variation of either signed muon flux is constrained,
and when the target neutrino is muon neutrino ($\nu_\mu$ and $\bar\nu_\mu$),
the flux variations of  both signed muon fluxes are constrained.

\begin{figure}[ht]
\includegraphics[height=5.5cm]{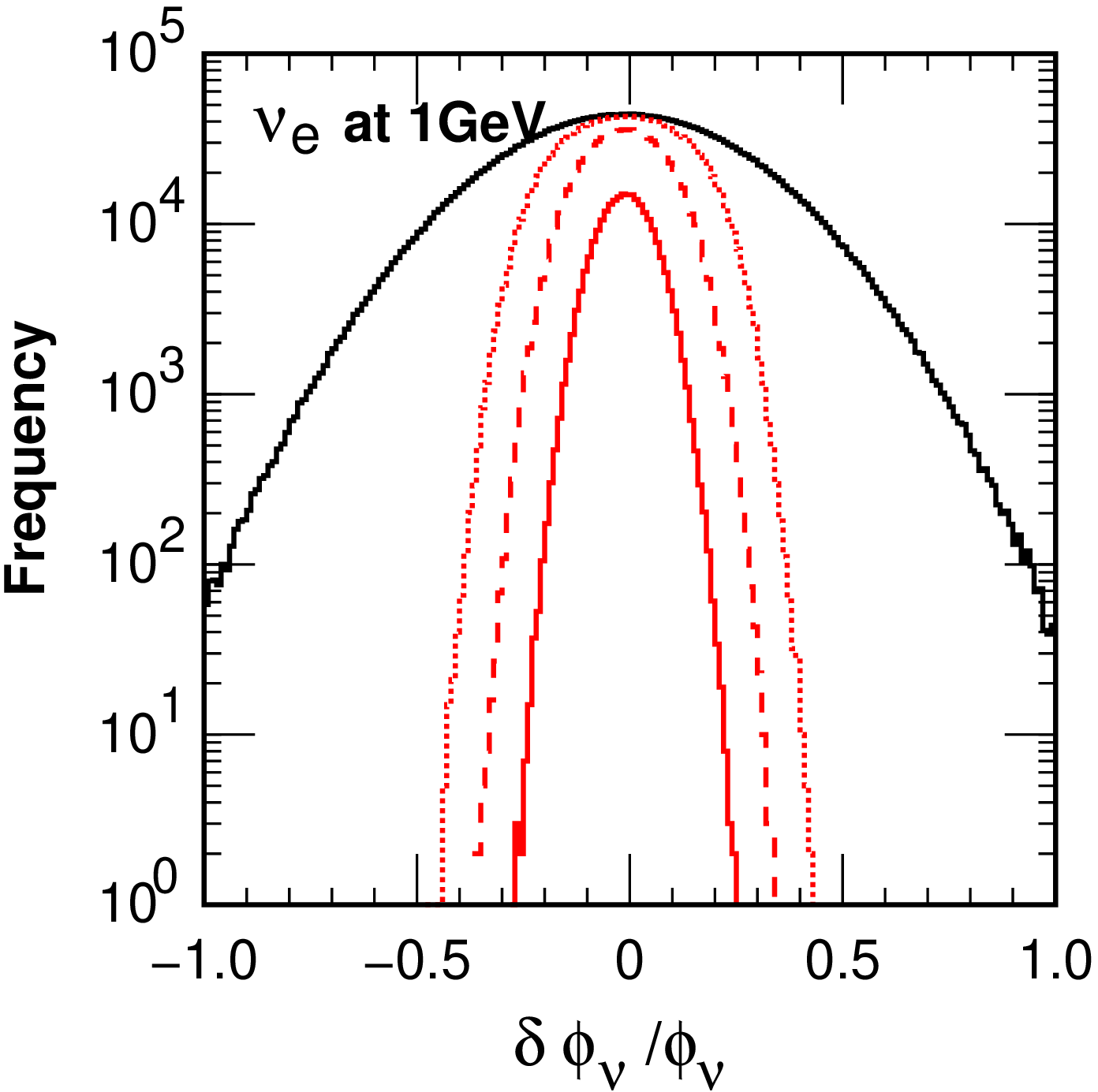}\hspace{5mm}%
\includegraphics[height=5.5cm]{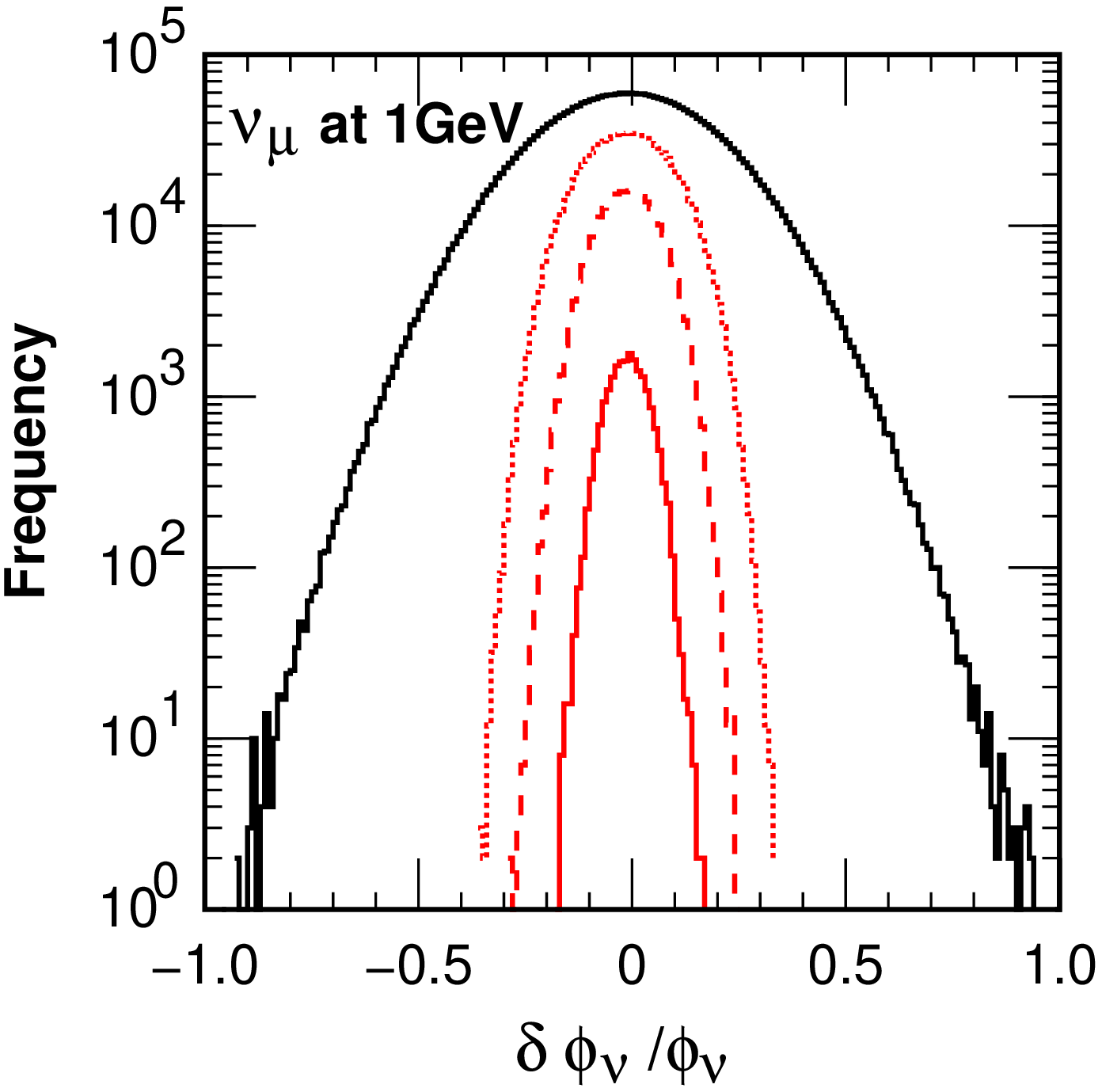}%
\caption{\label{varinu}
  The $\Delta\Phi_{\nu}/\Phi_{\nu}$ distributions for of $\nu_e$ flux (left panel) and
  $\nu_\mu$ flux (right panel), created with with  Eq.~\ref{bspl-df}  and $\delta=1$,
  with the 3,000,000 sets of normal random number $\{ R_N^{ij} \}$  assigned
  to each grid point.
  Wide outside solid line shows the distribution with no condition for
  $\Delta \Phi_\mu / \Phi_\mu$ ($\varepsilon=\infty$),
  most inside solid line for  $\Delta \Phi_\mu / \Phi_\mu <  0.1$  ($\varepsilon=0.1$),
  dashed line for  $\Delta \Phi_\mu / \Phi_\mu<  0.2$ ($\varepsilon=0.2$),
  and dotted line for
  $\Delta \Phi_\mu / \Phi_\mu<  0.3$ ($\varepsilon=0.3$).
  The integral kernels  we used are the
  vertically downward moving atmospheric neutrinos and muon fluxes
  at Kamioka.
    }
\end{figure}

In Fig.\ref{varinu},
we plot the variation of
$\Delta\Phi_{\nu}/\Phi_{\nu}$ at 1~GeV for $\nu_e$ in the left panel
and for $\nu_\mu$ in the right panel with  
$\varepsilon=$ 0.1, 0.2, 0.3,  and the ones without any constraint
($\varepsilon=\infty$).
In this plot, we used the integral kernel for vertically downward moving
atmospheric neutrino  observed at Kamioka,
and the integral kernel for  vertically downward moving muon fluxes
observed at Kamioka for the illustration.

We  find that  the distribution of  $\Delta\Phi_{\nu}/\Phi_{\nu}$  shrinks
in  both the panels with the decrease in $\varepsilon $.
 Considering the  interaction model we are using is a variation of the
ideal one which gives the real atmospheric neutrino
and the muon fluxes,
this observation could be interpreted as follows;
when our calculated atmospheric muon flux is close to the
real one, the atmospheric neutrino flux calculated with
the same interaction model must be close to the real one.
Furthermore if we consider that the atmospheric muon flux observed by
a  precision experiment is very close to the real one,
we can replace above sentence to;
when we can reconstruct the observed atmospheric
muon flux observed by a precision experiment,
the atmospheric neutrino flux calculated is close to the
real one.
Note, this arguments are  already discussed qualitatively in the other
article~\cite{shkkm2006},
but  this  variation study of the hadronic interaction model
gives a method for the quantitative discussion.

For each $\varepsilon$, the $\Delta\Phi_{\nu}/\Phi_{\nu}$ distribution
is  well approximated by the normal distribution as
\begin{equation}
  \rho\left(\varepsilon , {\Delta\Phi_{\nu} \over \Phi_{\nu}}\right) =
      {N_\varepsilon \over \sqrt{2\pi}}
      \exp\left({1 \over  \sigma_\varepsilon^2}
      \left( {\Delta\Phi_{\nu} \over \Phi_{\nu}}\right)^2 
      \right)\  ,
      \label{ndist}
\end{equation}
$N_\varepsilon$ is the total number of the trial which path the
limitation of the variation on the atmospheric muon flux.
We note  that the distribution without the constraint on the atmospheric
muon is also well approximated by the normal distribution then we we
use the same distribution formula Eq.~\ref{ndist} with
$\varepsilon=\infty$ and  $N_\infty$, which is the trial number of this study.
The concentration  of the neutrino flux variation distribution when
the variation of atmospheric muon flux is constrained with $\varepsilon$
may be
studied by the ratio of $ {\rho\left(\varepsilon , {\Delta\Phi_{\nu} / \Phi_{\nu}}\right)}$
and $\rho\left(\infty , {\Delta\Phi_{\nu}  / \Phi_{\nu}}\right)$, after the normalization
as,
\begin{equation}
 \left [{1 \over N_\varepsilon}
    {\rho\left(\varepsilon , {\Delta\Phi_{\nu} \over \Phi_{\nu}}\right)}
   \right] /\left[
    {1 \over  N_\infty}
      \rho\left(\infty , {\Delta\Phi_{\nu} \over \Phi_{\nu}}\right)
    \right]
    =
     \exp\left(\left({1 \over  \sigma_\varepsilon^2}-{1 \over  \sigma_\infty^2}\right) \cdot
      \left( {\Delta\Phi_{\nu} \over \Phi_{\nu}}\right)^2 
      \right)
      \label{conc0}
\end{equation}
Therefore, we define the concentration parameter $\sigma_{shrink}$ as
\begin{equation}
  {1 \over {\sigma_{shrink}^2}} = {1 \over {\sigma^2({\varepsilon })}}
  -  {1 \over {\sigma_{\infty}^2}}\ \ \ ,
      \label{shrink}
\end{equation}  
$\sigma_{shrink}$ would be the standard deviation of
the atmospheric neutrino flux variation, when the original
distribution of it  is flat, and the variation of atmospheric muon flux
 is restricted by Eq.~\ref{mulimit}.

 Let us consider the variation of atmospheric neutrino
 flux is the combination  of two components;
one is independent of that of atmospheric muon flux,
and the other is related to that of atmospheric muon flux.
When $\varepsilon$ approaches 0 in Eq.~\ref{mulimit},
the remaining variation of atmospheric neutrino flux would be
the component independent of the atmospheric muon flux.
We assume a simple function form for  $\varepsilon $ and
$\sigma_{shrink}$ as
\begin{equation}
  \sigma_{shrink}
  = \sqrt{
     \varsigma_{0}^2    + 
    \left(\varsigma_1 \cdot \varepsilon \right)^2} \ \ ,
\label{sigmafit}
\end{equation}  
where
$ \varsigma_0$ represents the atmospheric neutrino flux variation
independent of  the atmospheric neutrino flux,
and 
$\varsigma_1 \cdot  \varepsilon $ the atmospheric neutrino flux
variation related to the atmospheric muon flux.

Adding a little more data points, we fit the $\sigma_{shrink}$ 
with Eq.~\ref{sigmafit} for the atmospheric neutrino variation
distribution shown in Fig.~\ref{varinu},
and show the best fit curves for $\nu_e$ (left panel)
and $\nu_\mu$ (right panel).
We find Eq.~\ref{sigmafit} fit well both data,
and $ \sigma_{shrink}$ is already very close to the  $ \varsigma_0$
at $\varepsilon \sim 0.05$.
Note, we intend to apply this analysis to find a better interaction model
for the calculation of atmospheric neutrino flux
by the reconstruction test of the atmospheric muon flux observed by
a precision experiment.
Then $\Delta\phi_\mu / \phi_\mu \lesssim 0.05$ 
would be the practical target in this test.

\begin{figure}[ht]
\includegraphics[height=5.5cm]{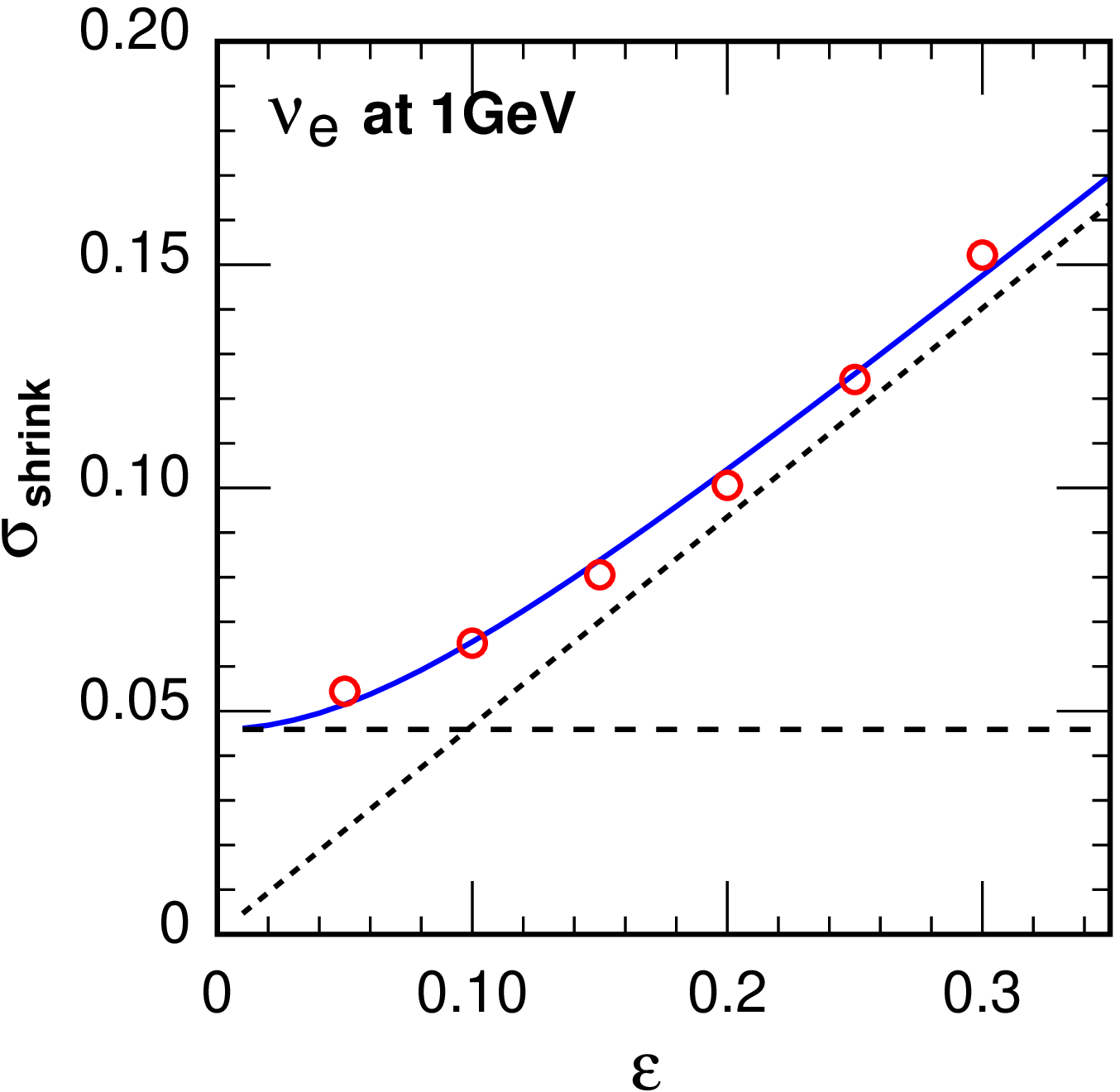}\hspace{5mm}%
\includegraphics[height=5.5cm]{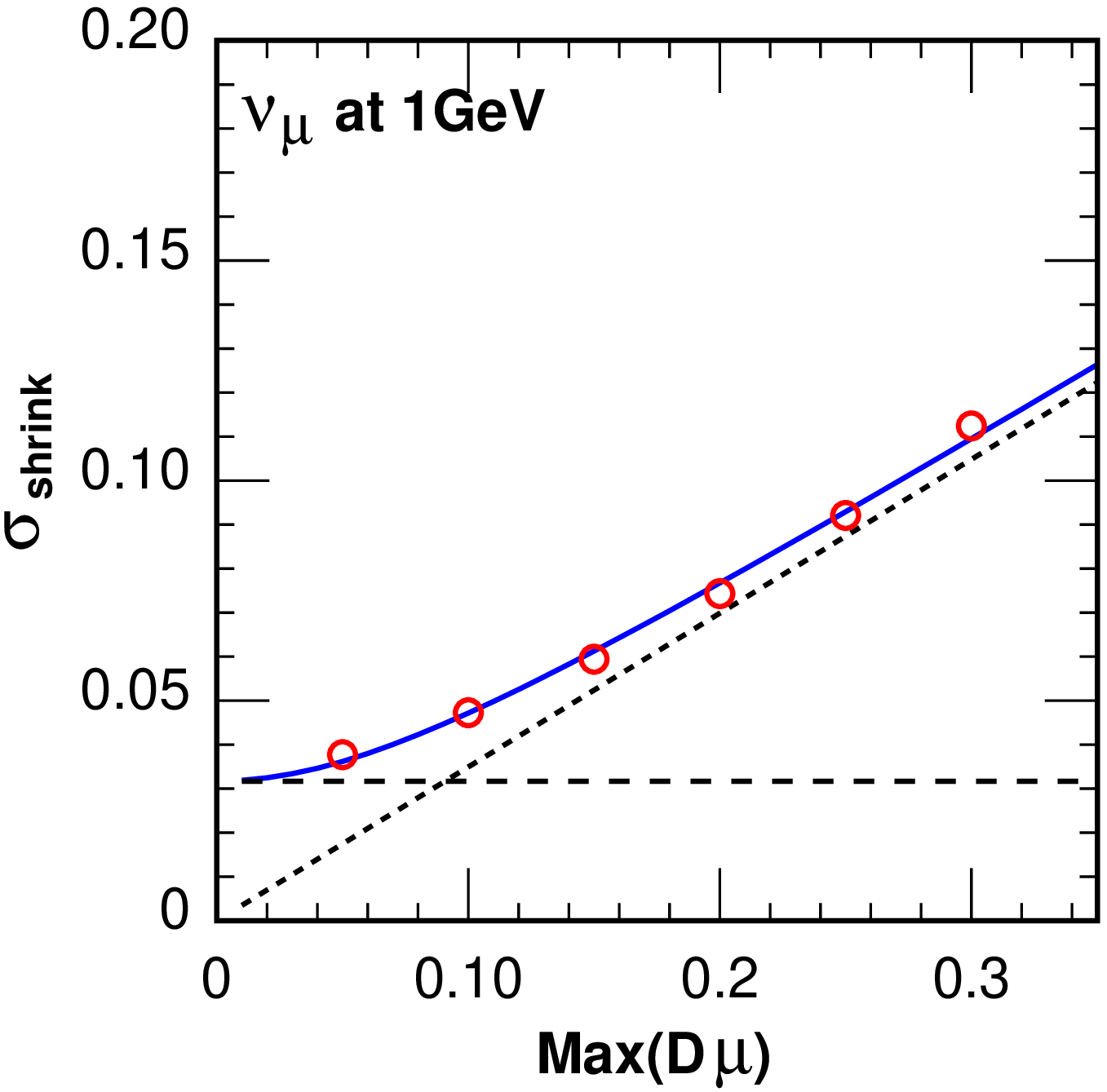}%
\caption{\label{merrfit}
  The  $\sigma_{shrink}$ obtained by the variation study of  atmospheric neutrino
  flux for vertically downward moving atmospheric electron neutrino 
  and downward moving atmospheric muon fluxes at Kamioka.
  for $\varepsilon$ = 0.5, 1.0, 1.5, 2.0, 2.5 and 3.0.
  The solid curve shows the best fit curve with Eq.~\ref{sigmafit}, and 
  the dash and dotted lines show two asymptotic lines of Eq.~\ref{sigmafit}; 
  $\sigma_{shrink}=\varsigma_0$, and 
  $\sigma_{shrink}=\varsigma_1 \cdot \varepsilon$.
We show those for $\nu_e$ in the left panel, and for $\nu_\mu$ in the right panel.
}
\end{figure}

\begin{figure}[ht]
\includegraphics[height=5.5cm]{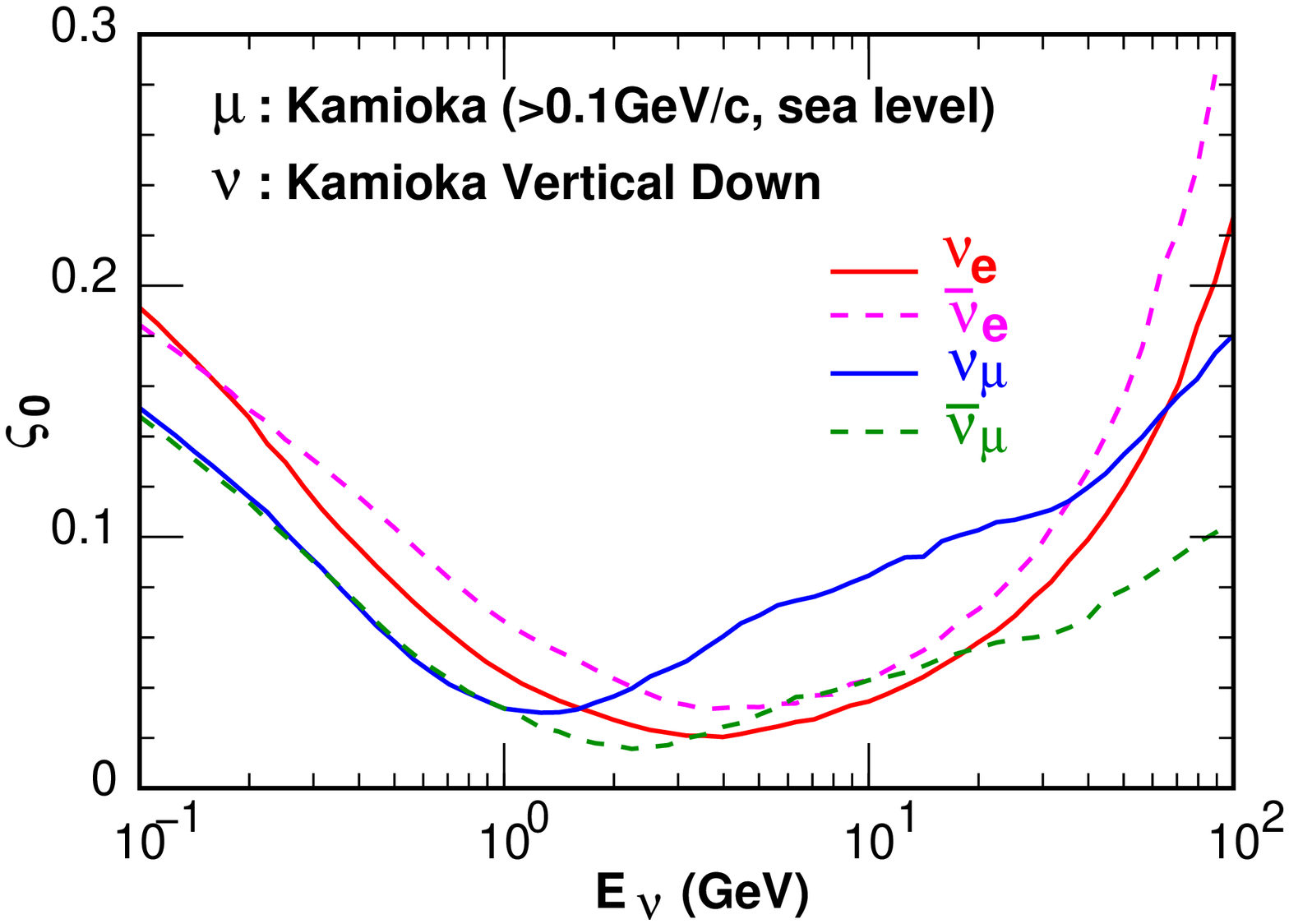}\hspace{5mm}%
\includegraphics[height=5.5cm]{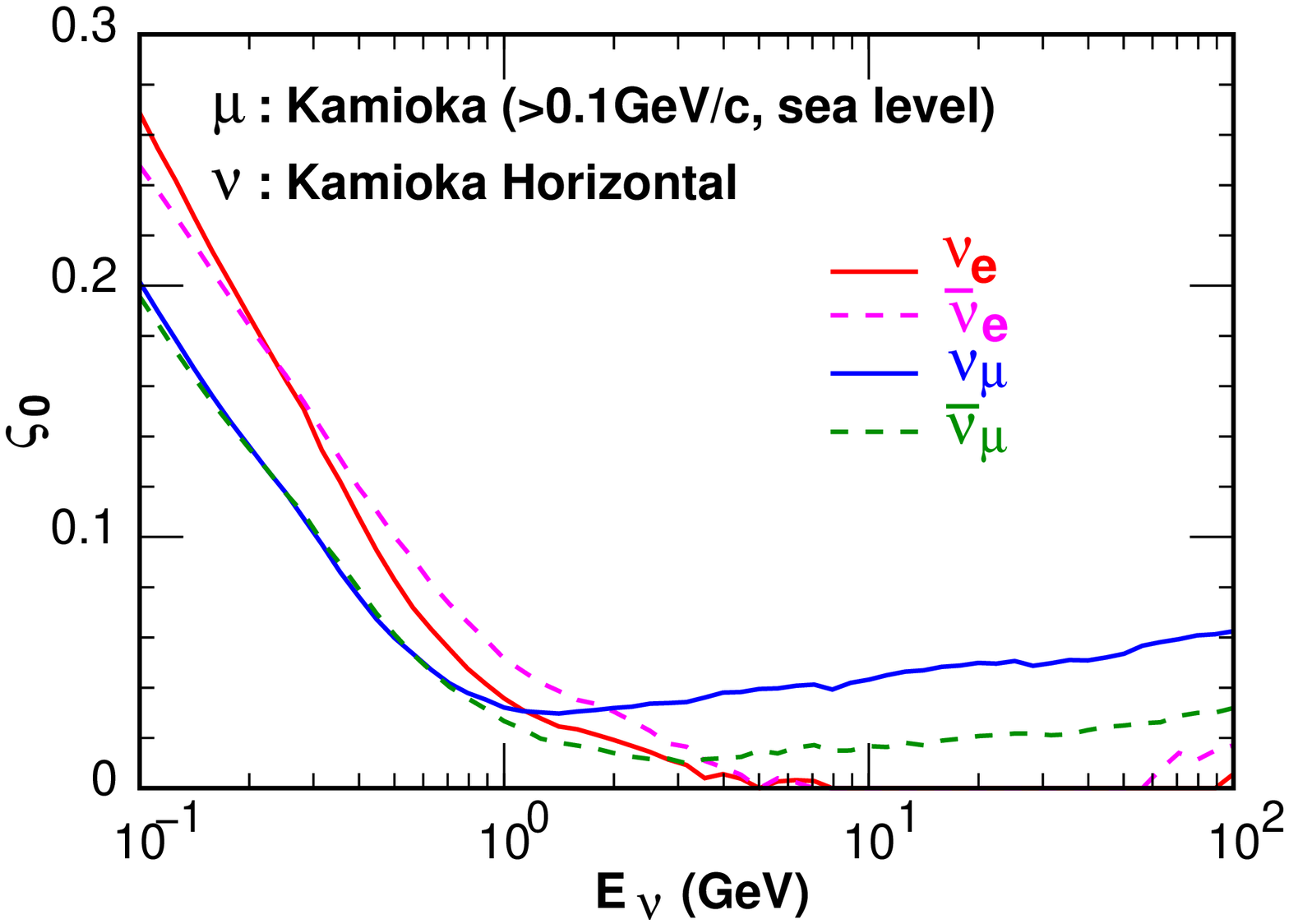}
\caption{\label{nerrkam01} 
  The $\varsigma_0$ or the atmospheric muon independent variation component of
  neutrino flux, calculated with the muon flux integral kernel for 
  the vertically downward moving  atmospheric muon flux at Kamioka (sea level).
  The minimum muon momentum is set to 0.1 GeV$/$c.
  In the left panel, we depicted the $\varsigma_0$ for vertical downward moving
  atmospheric neutrino at Kamioka, and in the right panel for horizontal moving
  atmospheric neutrino at Kamioka.
}
\end{figure}

Using the integral kernel for vertically downward moving atmospheric muon flux
at Kamioka,
we repeat the variation study with $\varepsilon$ = 0.5, 1.0, 1.5, 2.0, 2.5 and 3.0,
and fit the resulting $\sigma_{shrink}$  by  Eq.~\ref{sigmafit}  to determine
 $\varsigma_0$ or the muon independent component of
the atmospheric neutrino flux variation,
for vertically downward and horizontally moving atmospheric
neutrino fluxes at Kamioka in the energy range 
from 0.1 GeV to 100 GeV beyond our target,
and for all kind of neutrinos.
We plot the $\varsigma_{0}$
in Fig.~\ref{nerrkam01}  as the function of neutrino energy.
Note, we set the minimum muon momentum for this study to 0.1 GeV$/$c.
This means that
we study the correlation coefficient of neutrino and muon fluxes
in the muon momentum range larger than 0.1 GeV$/$c for a given neutrino energy.
Then we constrain the muon flux variation at the muon momentum 
where the correlation coefficient is larger than the 90~\% of the maximum.

The most crucial fact in Fig.~\ref{nerrkam01}  is that
$\varsigma_0$ increases  rapidly as the neutrino energy decrease
below 1 GeV, for all the kind of neutrinos.
In the next section, we will discuss on the rise of  $\varsigma_0$ at
low energies.
Although the energy region is out of our target energy region,
we have some comments on the $\varsigma_0$ increase with neutrino energy 
above a few GeV.
This is due to the kaon contribution to neutrino production, whose
variation is not 
restricted by the limitation of the variation of the atmospheric muon flux.
As the kaon contribution is largest to $\nu_\mu$ production among
all the neutrinos,  the $\varsigma_0$ increase of $\nu_\mu$ is largest among them.
Note, we have also assumed the uncertainty 
of kaon production is 50~\% at the every grid point
of the integral kernel in Eq.~\ref{bspl-df}.
If we apply here the uncertainty of kaon production by accelerator
experiment, the increase of  $\varsigma_0$ would be suppressed.
For the horizontally moving neutrinos, still the increase of $\varsigma_0$
is seen for  $\nu_\mu$, but generally it stays $ \lesssim 0.05$ below
100~GeV. 

\section{\label{olddata} analysis of existing data}

\begin{figure}[ht]
  \includegraphics[height=5.5cm]{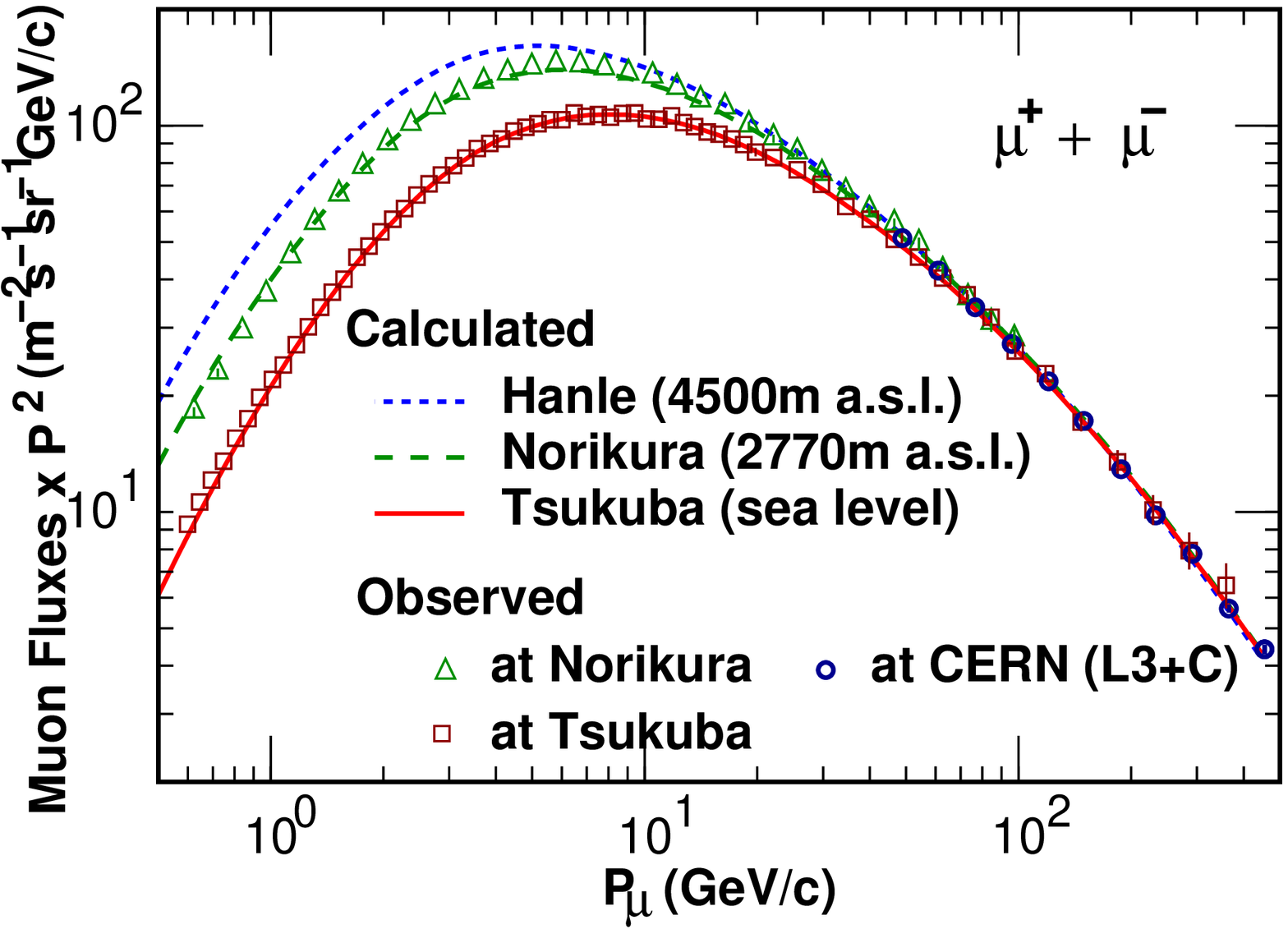}\hspace{5mm}%
  \includegraphics[height=5.5cm]{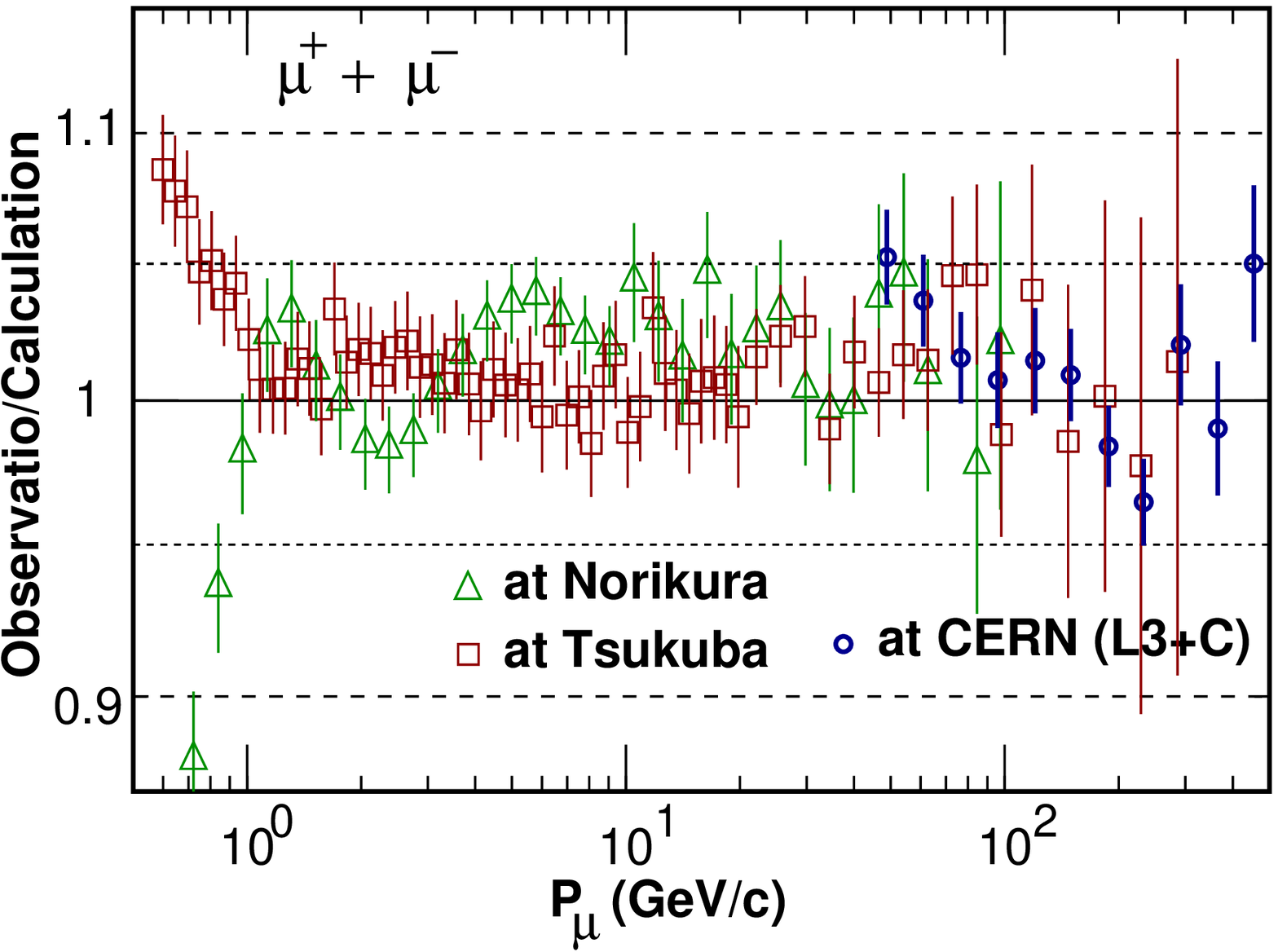}
  \caption{\label{mudata}
    Left panel: vertically downward atmospheric muon flux 
    ($\mu^+ + \mu^-$ ) observed  by the  BESS detector 
    at Tsukuba (sea level, squares)~\cite{BESSTeV}
    and at  M. Norikura (2770m a.s.l., upper triangles)~\cite{BESSnorimu},
    by L3+C experiment~\cite{l3+c} at CERN (circles).
    Solid line is the vertically downward  atmospheric
    muon flux calculated for Tsukuba,
    dashed line  for Mt. Norikura,
    dotted line for Hanle (4500m a.s.l.  India).
    Right panel: Expanded comparison of observation and calculation
    taking their ratio.
}
\end{figure}

In our former study~\cite{shkkm2006}, we have estimated our calculation
error for the atmospheric neutrino flux using the atmospheric muon spectra 
observed by the BESS detector at Tsukuba (sea level)~\cite{BESSTeV},
at Mt. Norikura (2770m a.s.l.)~\cite{BESSnorimu} above 0.567 GeV$/$c.
In the left panel of Fig.~\ref{mudata}, 
we plot these data taking the flux sum of $\mu^+$ and $\mu^-$ 
with the data observed by L3+C experiment~\cite{l3+c} at CERN.
We also depict the calculated fluxes for these observation sites,
and for Hanle (India, 4500m a.s.l.) in the same figure.
In the right panel of Fig.~\ref{mudata}, we show the comparison of
observed and calculated atmospheric muon flux expanding 
the difference  by taking the ratio.
Note,
we have renewed the calculation of the muon flux with the
primary cosmic ray model based on 
AMS02 and other precision measurements
\cite{ams02p, ams02he, pamela2013, besspolar}.

\begin{figure}[ht]
\includegraphics[height=5.5cm]{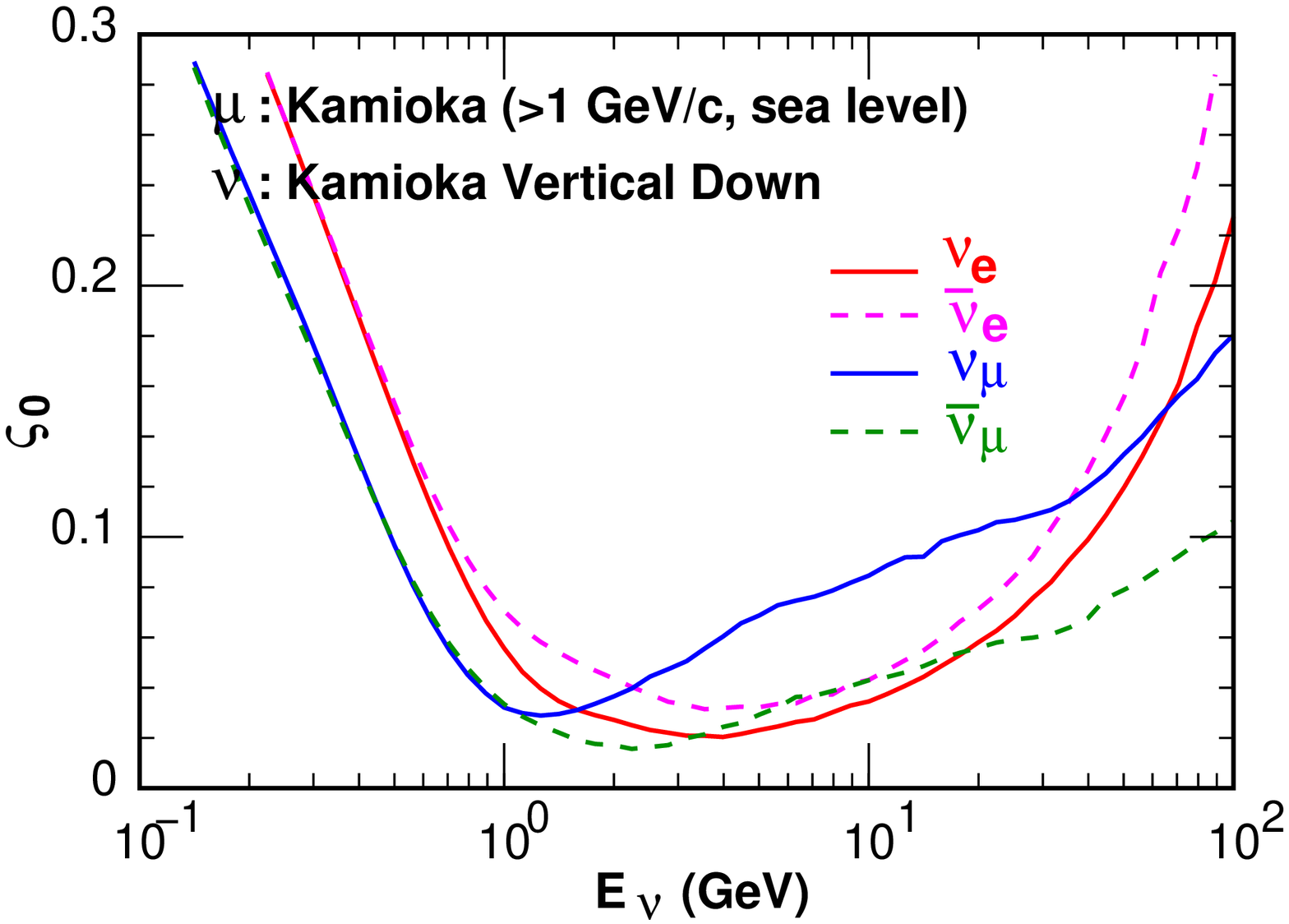}\hspace{5mm}%
\includegraphics[height=5.5cm]{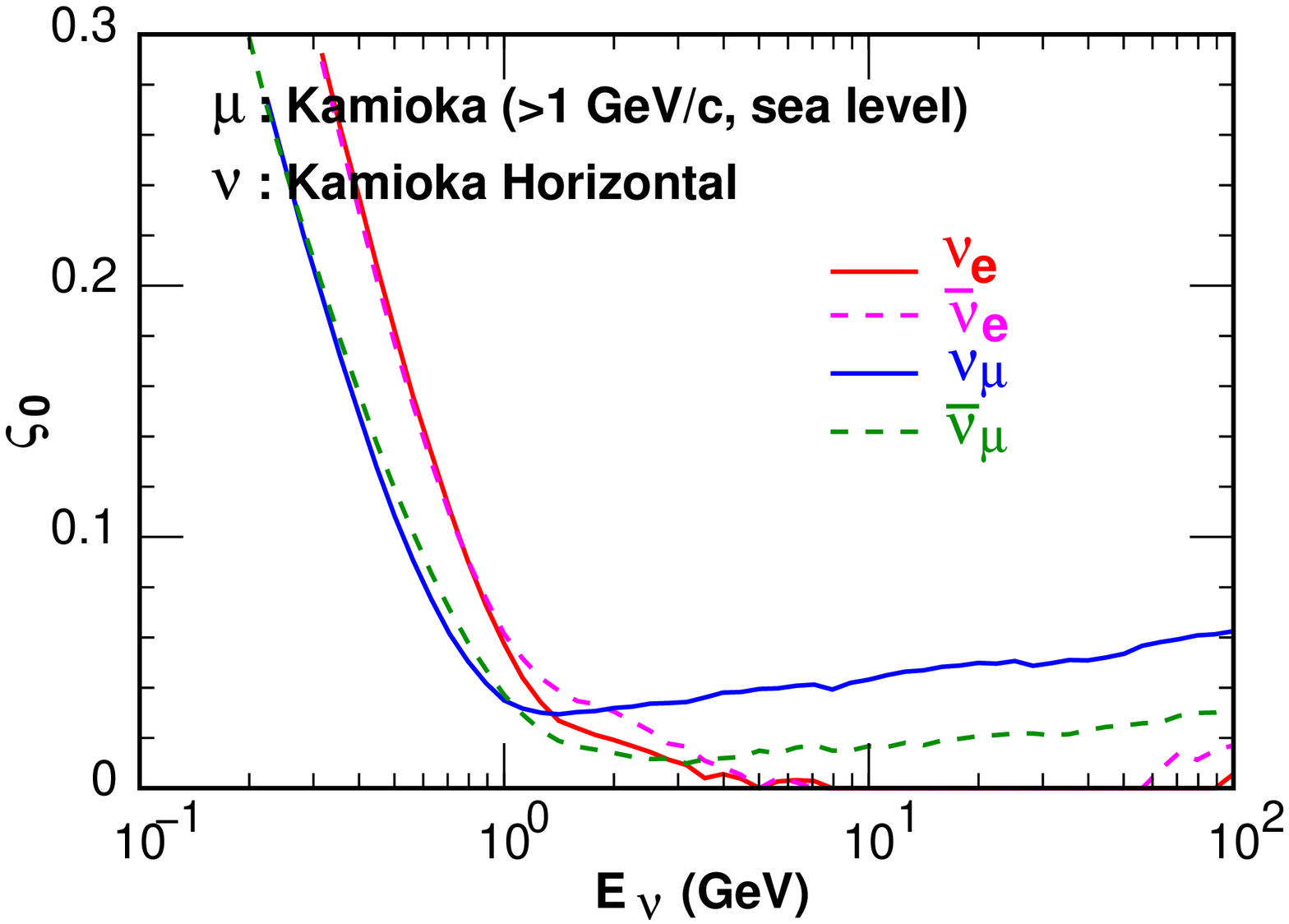}
\caption{\label{nerrkam10} 
  The $\varsigma_0$ or the atmospheric muon independent variation component of
  neutrino flux, calculated with the muon flux integral kernel for 
  the vertically downward moving  atmospheric muon flux at
  Kamioka (sea level).
  The minimum muon  momentum is set to 1 GeV$/$c.
  In the left panel, we depicted the  $\varsigma_0$ for vertical downward moving
  atmospheric neutrino at Kamioka, and in the right panel for horizontal moving
  atmospheric neutrino at Kamioka.
  }
\end{figure}

We apply the study in the previous section to these data, especially to
those observed by BESS.
Note, 
the agreement of calculation and the observed data for atmospheric muon
flux is generally
good and the reconstruction residual is less than 5~\% above 1 GeV.
However,  below 1~GeV, we failed to reconstruct the observed muon flux
at Tsukuba and at Mt. Norikura at the same time.
Probably we need new observation by a precision experiment,
dedicated to the low momentum muon flux ($\lesssim$ 1 GeV).
In the previous section, we calculate  $\varsigma_0$ with the minimum
muon momentum of 0.1 GeV$/$c (Fig.~\ref{nerrkam01}),
implicitly assuming that we can reconstruct the the accurately measured
muon flux for $P_\mu \gtrsim$ 0.1 GeV$/$c.
In the previous section,
we have calculate  the $\varsigma_0$ with the integral kernel for the
vertically downward moving atmospheric muon flux at Kamioka,
where the muon observation condition is very close to Tsukuba,
and here we re-calculate it and plot in  Fig.~\ref{nerrkam10} with a little change
that the minimum muon momentum is set to be 1 GeV$/$c.
Note, we plot the $\varsigma_0$ rather than the $\sigma_{shrink}$ with
the residual shown in the left panel of  Fig.~\ref{mudata},
since the estimation of residual at each momentum is difficult
due to the muon flux observation error,
but it would be smaller than 0.05 in Fig.~\ref{mudata}.
Note, 
$\sigma_{shrink}$ at $\varepsilon$=0.05 is very close to 
$\varsigma_0$.

The $\varsigma_0$ or the atmospheric muon independent variation component of
  neutrino flux  in Fig.~\ref{nerrkam10} is compared with that
in Fig.~\ref{nerrkam01}, and we find the $\varsigma_0$ in Fig.~\ref{nerrkam10}
shows a quicker increase towards lower energy below 1 GeV,
but it is very similar in each figure above $\sim$ 1 GeV.
We may say that, using the atmospheric muon data observed by BESS at Mt. Norikura,
we can draw almost the same conclusion on the uncertainty of the low
energy atmospheric neutrino flux as that in Ref.~\cite{shkkm2006}.

\section{\label{futuredata}Survey of muon observation site  for future experiments}

In the previous section we applied the study in Sec.~\ref{variflx} to the
presently available data. We could confirm the result of our former study,
but also find we need more muon flux data with a precision experiment
dedicated to lower muon momentum.
However, the comparison of Fig. ~\ref{nerrkam01} and Fig.~\ref{nerrkam10},
tells us that,
just by lowering the minimum muon momentum, it is difficult to reduce the
uncertainty of low energy atmospheric neutrino flux largely with the muon
observation at sea level.
Then we look for suitable muon observation site for the future atmospheric
muon observation experiment.
Considering the progress of the detectors used in the recent  cosmic ray o
observations,
we assume that we can get the accurate muon flux data from 0.3 GeV$/$c
in a precision experiment for the low energy muon flux observation.

\begin{figure}[ht]
\includegraphics[height=5.5cm]{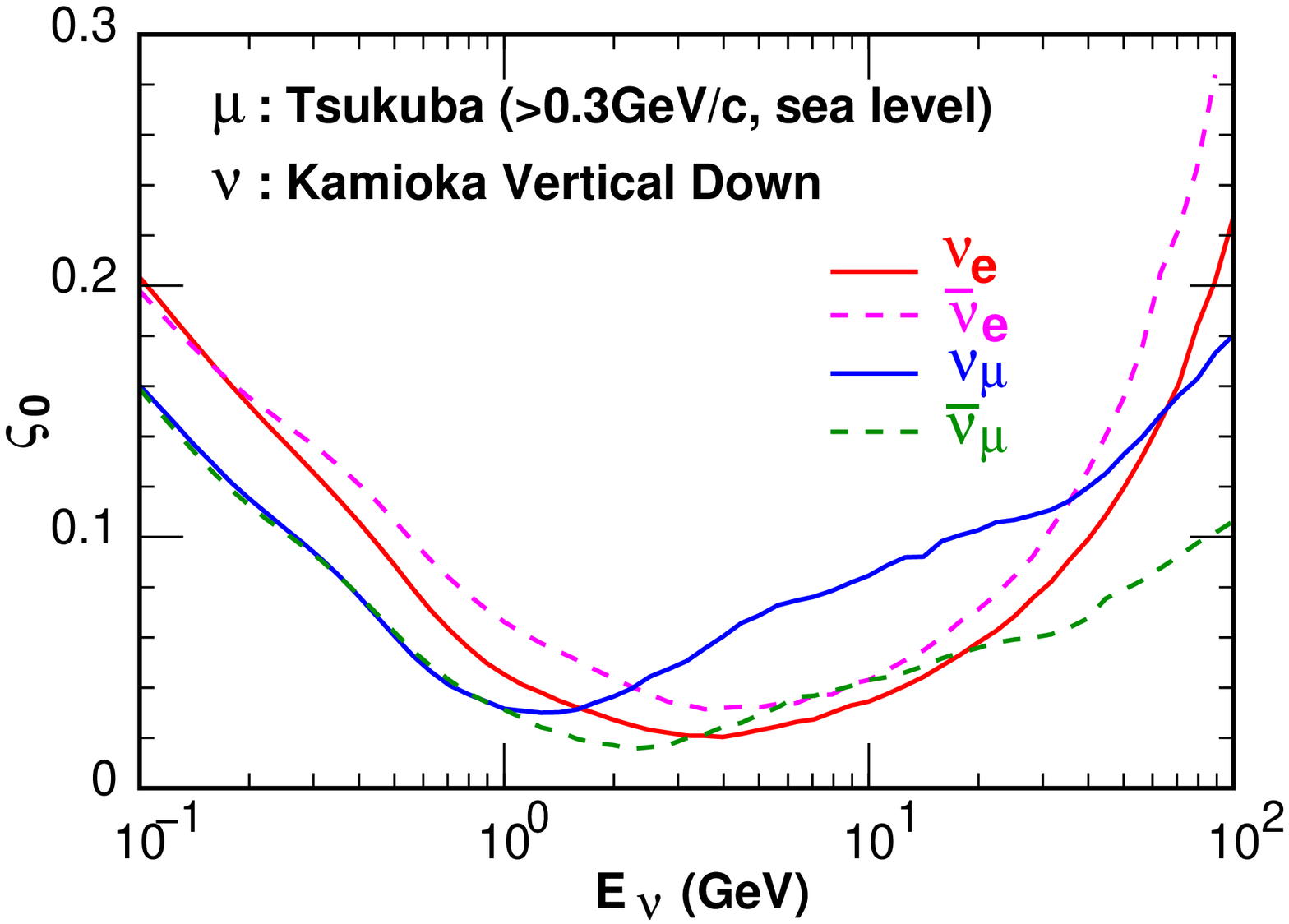}\hspace{5mm}%
\includegraphics[height=5.5cm]{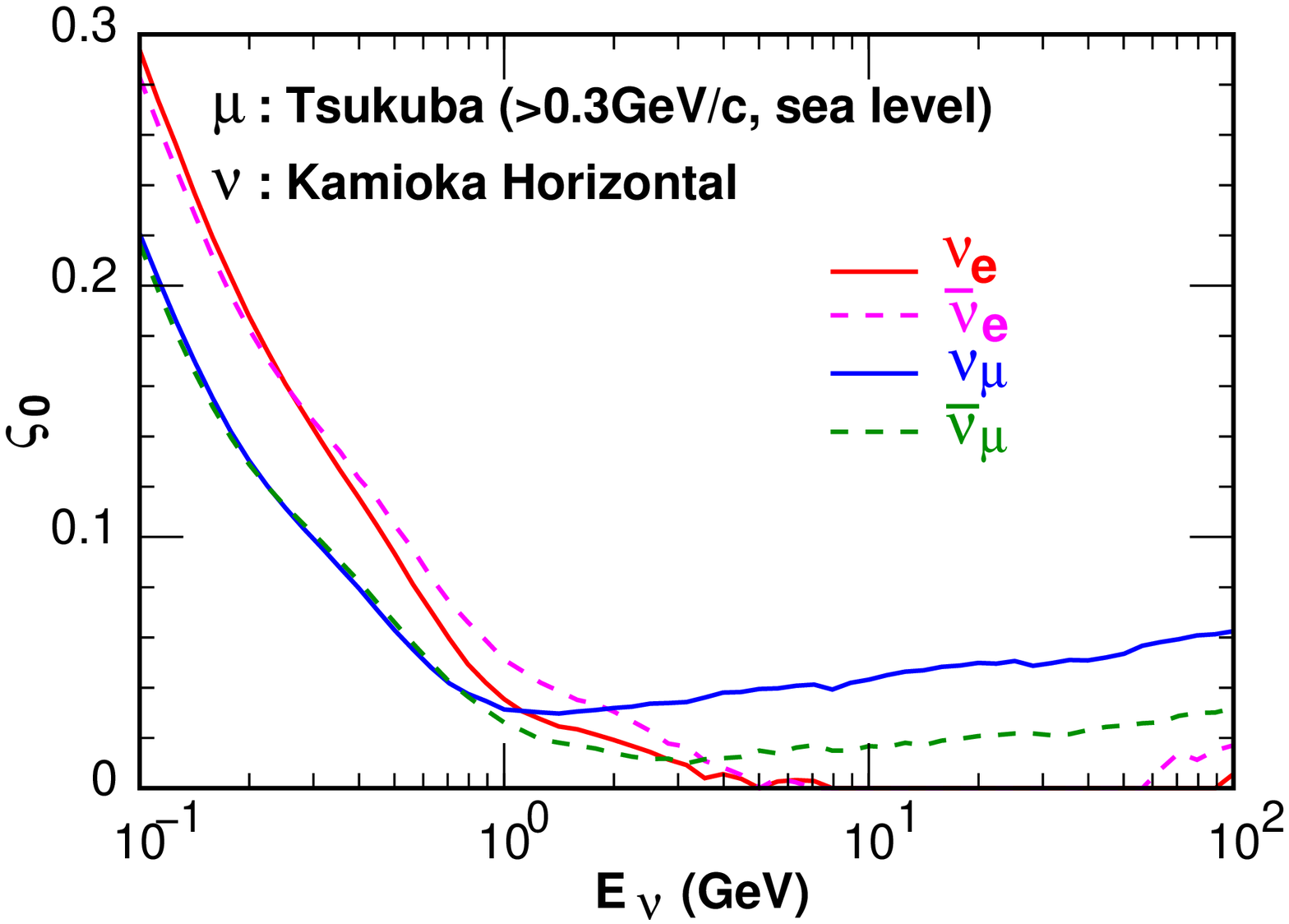}
\caption{\label{nerrtsuku03} 
  The $\varsigma_0$ or the atmospheric muon independent variation component of
  neutrino flux, calculated with the muon flux integral kernel for 
  the vertically downward moving  atmospheric muon flux at
  Tsukuba (sea level).
  The minimum muon momentum is set to 0.3 GeV$/$c.
  In the left panel, we depicted the $\varsigma_0$ for vertical downward moving
  atmospheric neutrino at Kamioka, and in the right panel for horizontal moving
  atmospheric neutrino at Kamioka.
   }
\end{figure}

Let us start the survey from Tsukuba (sea level), where BESS group has observed the
muon flux as we stated in the previous section.
In Fig. ~\ref{nerrtsuku03}, we plot  $\varsigma_0$ for  
the  atmospheric neutrino flux at Kamioka, 
with the
integral kernel for the vertical downward moving atmospheric muon flux at Tsukuba
(sea level).
Since the observation altitude and the rigidity cutoff are very close to those at Kamioka,
we can compare this result  with those presented and discussed in the previous section. 
We find the result here is in between of those with the minimum muon momentum
of 0.1 GeV$/$c (Fig.~\ref{nerrkam01}) and of 1 GeV$/$c (Fig.~\ref{nerrkam10}),
and is rather close to the calculation with minimum momentum of 0.1 GeV$/$c.
Therefore, if the atmospheric muon flux is measured down to 0.3 GeV$/$c,
by a precision experiment, it will improve the result with former BESS
observation and constrain the uncertainty in the
atmospheric neutrino flux down to a little less than 1 GeV.
 
Next we move to Mt. Norikura (2770m a.s.l.), 
where BESS group also has observed the muon flux.
In Fig. ~\ref{nerrtsuku03}, we plot  $\varsigma_0$ for
the  atmospheric neutrino flux at Kamioka, 
with the
integral kernel for the vertical downward moving atmospheric muon flux at Mt. Norikura
(2770m a.s.l.l), and the minimum muon moment of 0.3~GeV$/$c.
We find the  $\varsigma_0$  at Mt. Norikura is similar to that at Tsukuba for $E_\nu > 0.5$~GeV, 
but show a large reduction for $E_\nu <  0.5$~GeV.
It is remarkable that  $\varsigma_0 < 0.1 $ is satisfied for each kind of neutrino in
0.3~GeV $< E_\nu <$ 10~GeV for vertical direction and in $E_\nu > 0.3$~GeV for
horizontal direction.
Thus the observation at the high altitude is seems to have an advantage in the reduction of
the uncertainty of the low energy neutrino flux prediction.

\begin{figure}[ht]
\includegraphics[height=5.5cm]{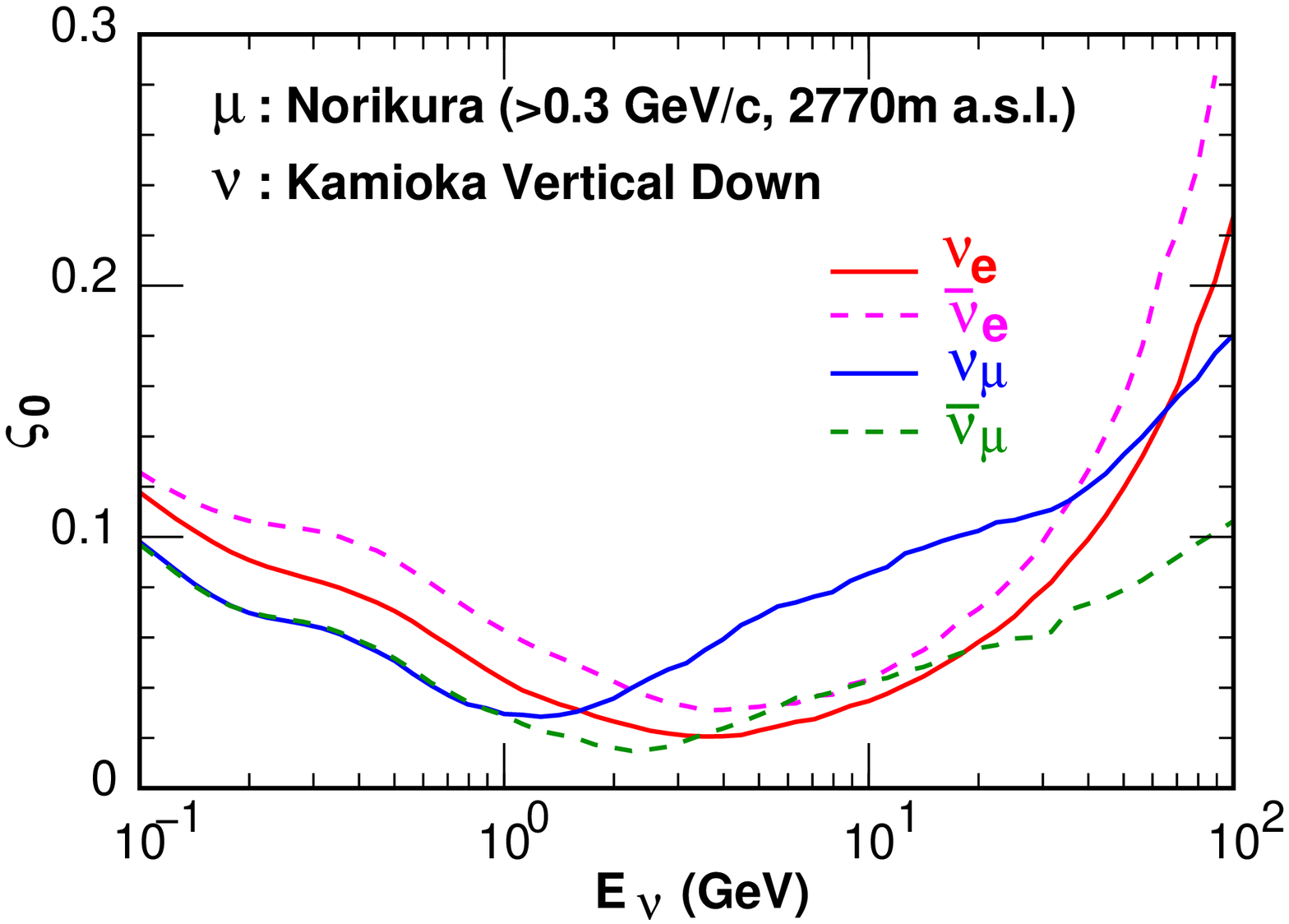}\hspace{5mm}%
\includegraphics[height=5.5cm]{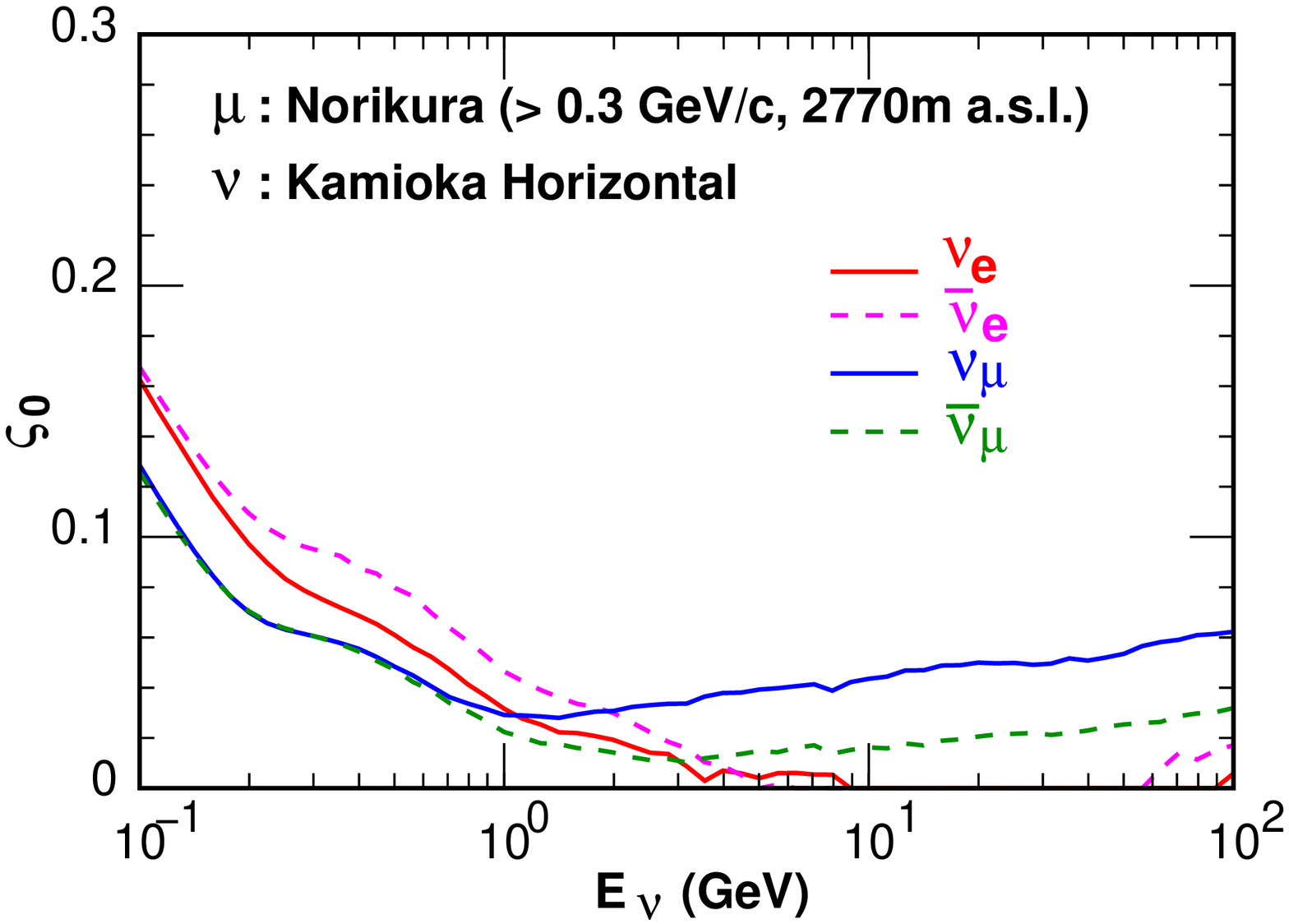}
\caption{\label{nerrnori03} 
  The $\varsigma_0$ or the atmospheric muon independent variation component of
  neutrino flux, calculated with the muon flux integral kernel for 
  the vertically downward moving  atmospheric muon flux at
Mt. Norikura (2770m a.s.l.).
  The minimum muon momentum is set to 0.3 GeV$/$c.
  In the left panel, we depicted the $\varsigma_0$ for vertical downward moving
  atmospheric neutrino at Kamioka, and in the right panel for horizontal moving
  atmospheric neutrino at Kamioka.
   }
\end{figure}

\begin{figure}[ht]
\includegraphics[height=5.5cm]{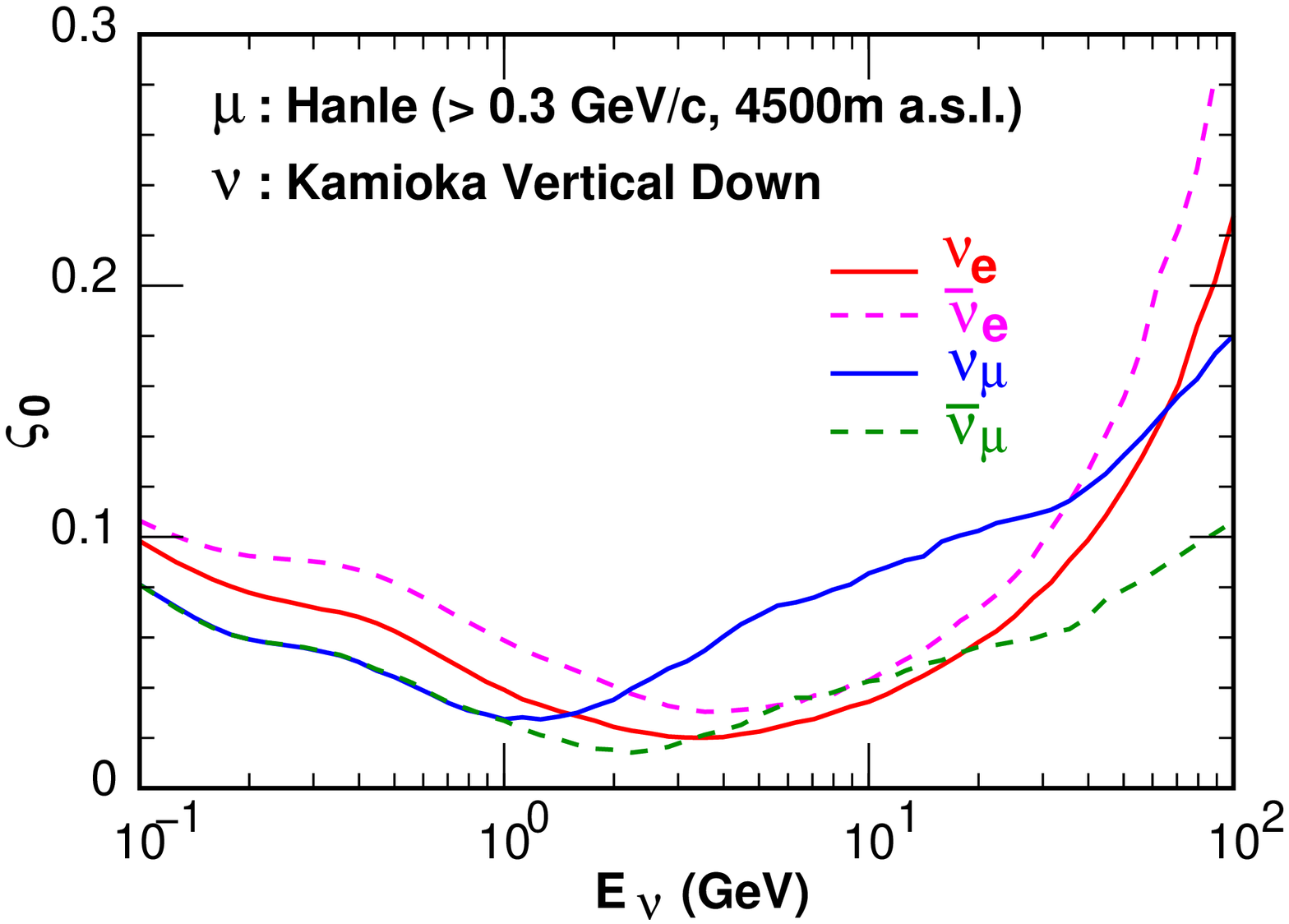}\hspace{5mm}%
\includegraphics[height=5.5cm]{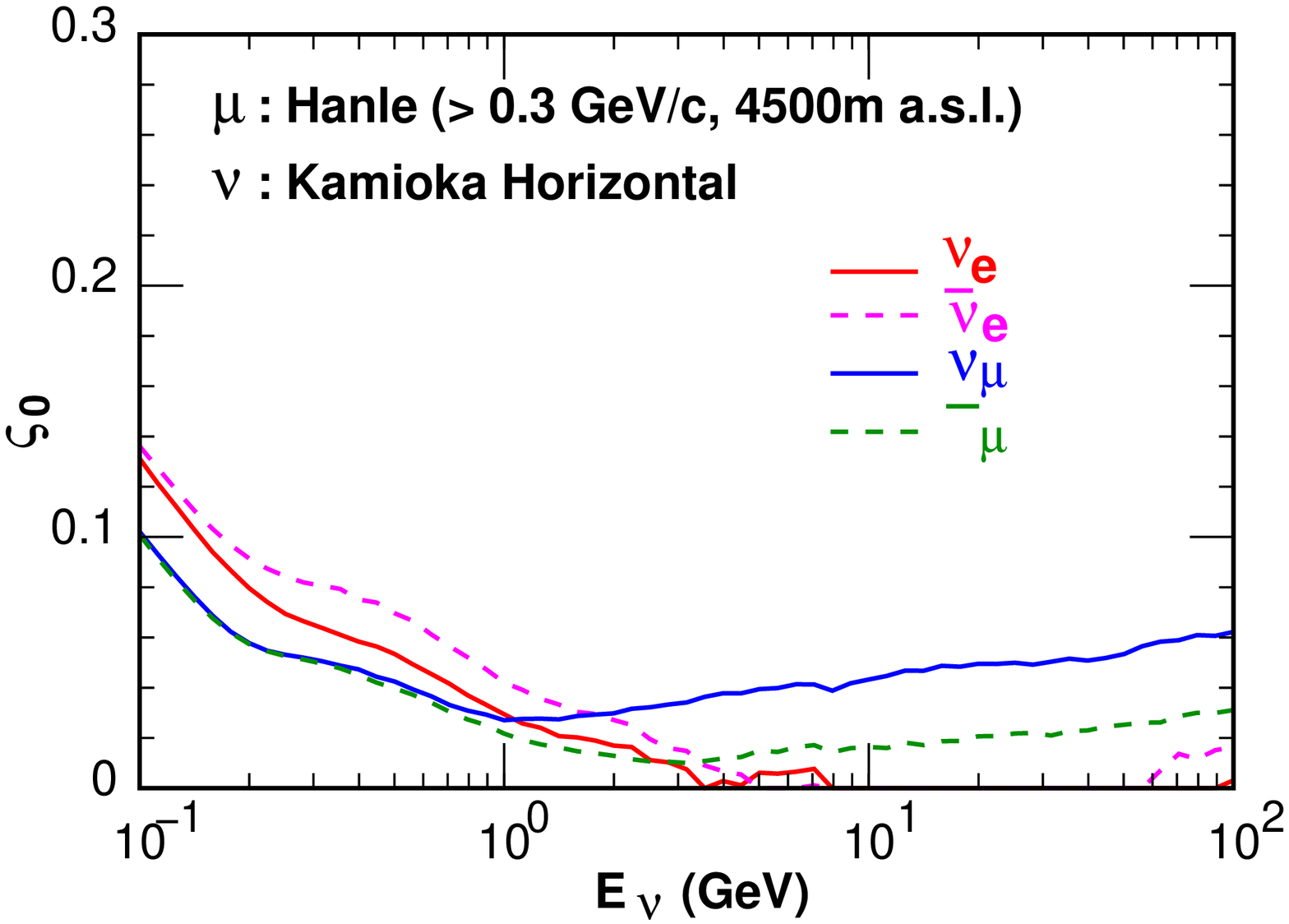}
\caption{\label{nerrHL03}  
  The $\varsigma_0$ or the atmospheric muon independent variation component of
  neutrino flux, calculated with the muon flux integral kernel for 
  the vertically downward moving  atmospheric muon flux at
  Hanle (4500m a.s.l.).
  The minimum muon momentum is set to 0.3 GeV$/$c.
  In the left panel, we depicted the $\varsigma_0$ for vertical downward moving
  atmospheric neutrino at Kamioka, and in the right panel for horizontal moving
  atmospheric neutrino at Kamioka.
  }
\end{figure}

As it seems the higher altitude is more suitable for the muon observation site,
we look for a candidate  at much higher altitude, and find
Hanle  (4500m a.s.l., India)~\cite{hanle} satisfies the condition,
where the Indian astronomical observatory exists.
We calculate the  $\varsigma_0$ for the atmospheric neutrino flux at Kamioka,
with the
integral kernel for vertical downward moving atmospheric muon at Hanle,
and plot it  In Fig. ~\ref{nerrHL03} .
The minimum muon momentum is set to 0.3 GeV$/$c as before.
Comparing with the  $\varsigma_0$ calculated with the muon at Mt. Norikura,
we find the  the  $\varsigma_0$ for Hanle is generally
smaller in $E_\nu< 1$ GeV.
Especially  $\varsigma_0< 0.1 $ is satisfied in $0.15 < E_\nu < 10$~GeV
for all the kind of neutrino and for all the directions.

\begin{figure}[ht]
  \includegraphics[height=5.5cm]{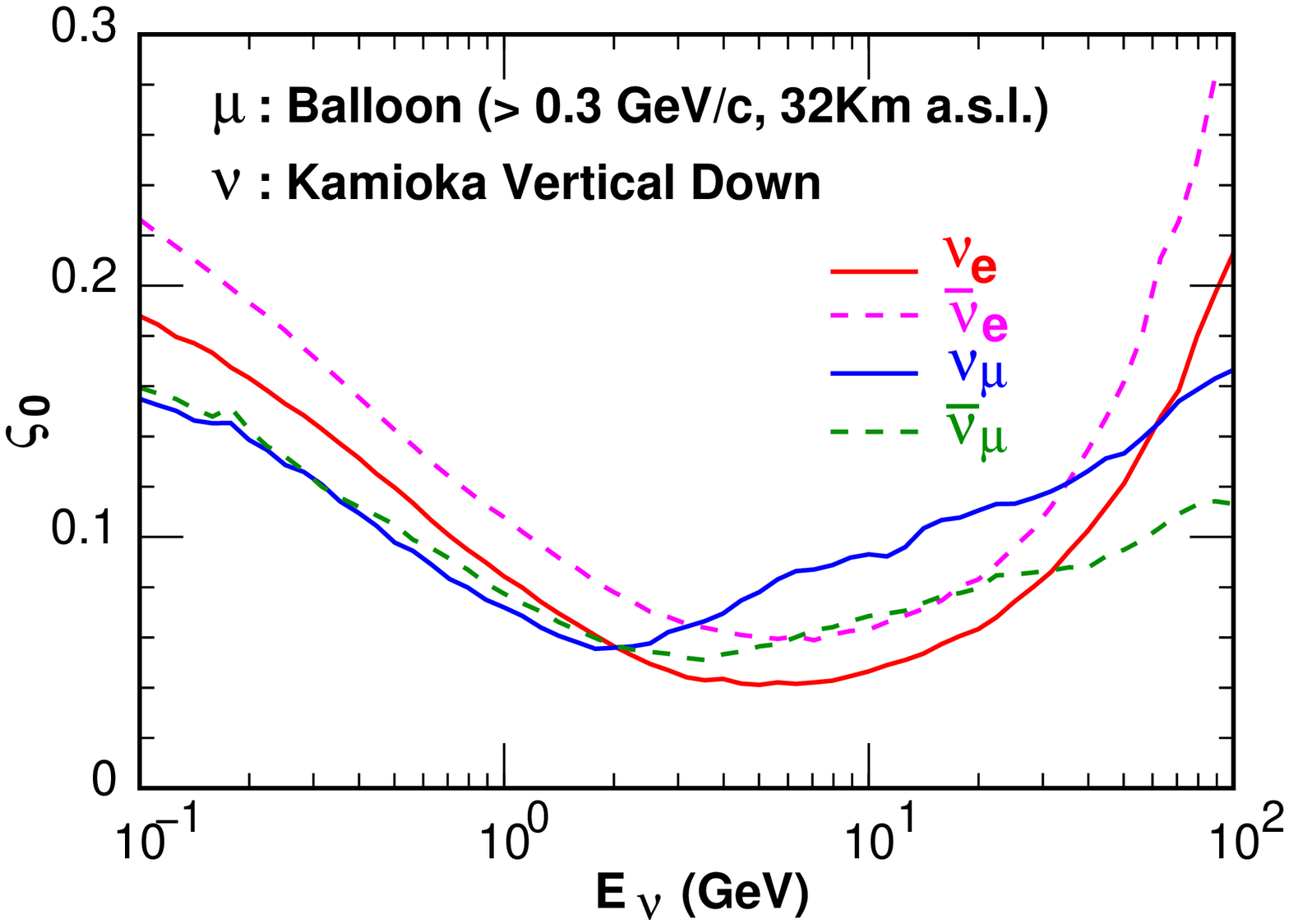}\hspace{5mm}%
  \includegraphics[height=5.5cm]{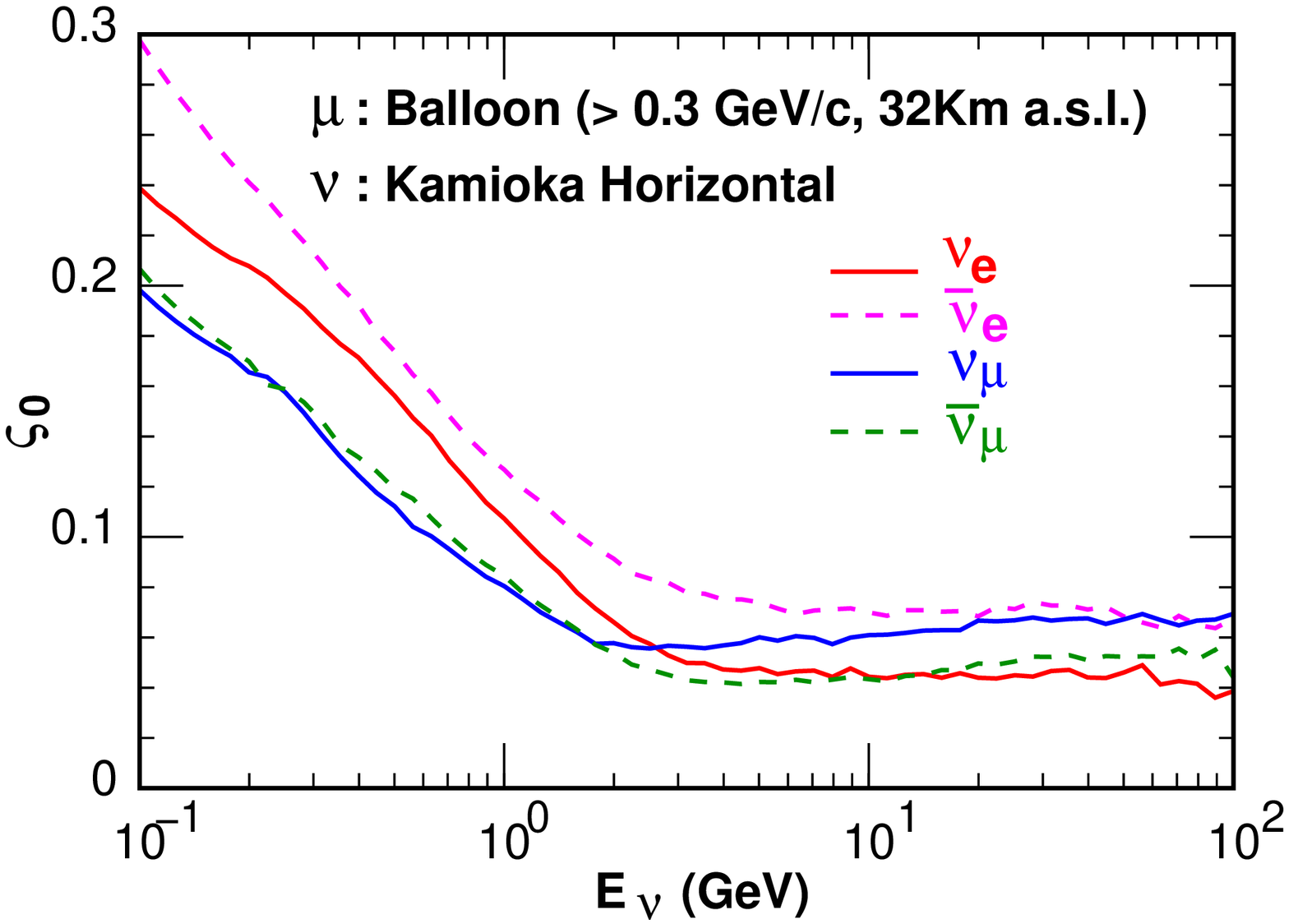}
  \caption{\label{nerrBP03}
  The $\varsigma_0$ or the atmospheric muon independent variation component of
  neutrino flux, calculated with the muon flux integral kernel for 
  the vertically downward moving  atmospheric muon flux at
  Balloon altitude (32km a.s.l.).
  The minimum muon momentum is set to 0.3 GeV$/$c.
  In the left panel, we depicted the $\varsigma_0$ for vertical downward moving
  atmospheric neutrino at Kamioka, and in the right panel for horizontal moving
  atmospheric neutrino at Kamioka.}
\end{figure}

The last candidate for muon observation site is the balloon which is used for the
observation of primary cosmic rays.
Note,
we have once used the balloon altitude muon  data observed by BESS~\cite{Abe:2003cd}
to study the interaction model at low energies~\cite{hkkm2011}.
However,  we could not conclude a strong statement due to the poor statistics.
We calculate the  $\varsigma_0$  for the atmospheric neutrino flux
at Kamioka with the
integral kernel for the vertical downward moving atmospheric muon at Balloon altitude
(32km a.s.l., near south pole),
and plot it in Fig. ~\ref{nerrBP03}.
Note, as the atmospheric muon flux at balloon altitude is small than the lower altitudes,
we consider a long flight balloon experiment for the observation of it.
The minimum muon momentum is set 0.3~GeV$/$c as before.
We find that the value of  $\varsigma_0$
is larger than those of others in the all neutrino energy region we studied.
This means the muon observation at balloon altitude does not reduce the
uncertainty of the low energy atmospheric neutrino flux more than that on a high
mountain. We have to look for a  good muon observation site on a high mountain.

\section{\label{other}
  Uncertainties of the Projectile flux and the scattering angle}

\begin{figure}[tbh]
\includegraphics[height=5.5cm]{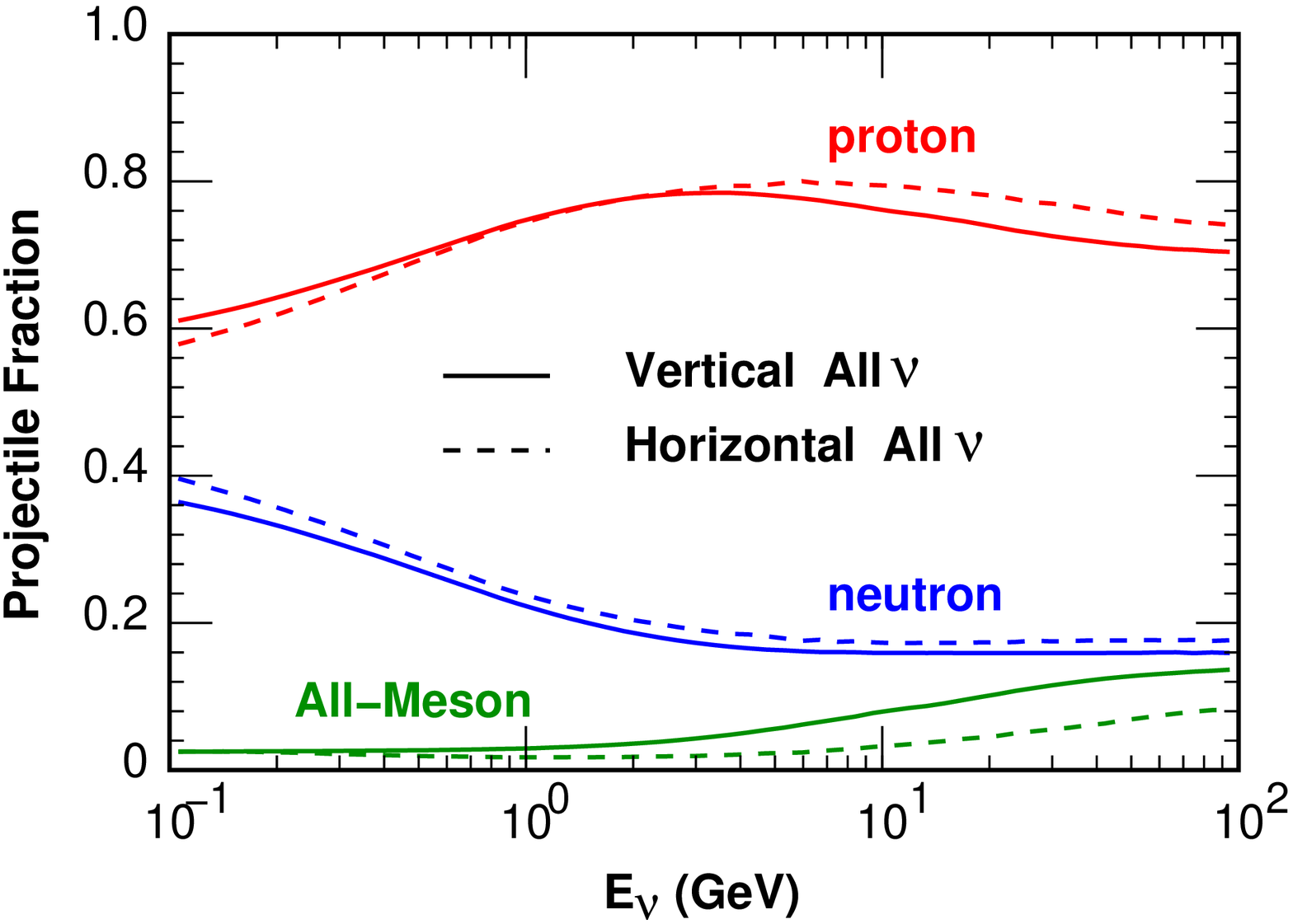}%
\hspace{5mm}
\includegraphics[height=5.5cm]{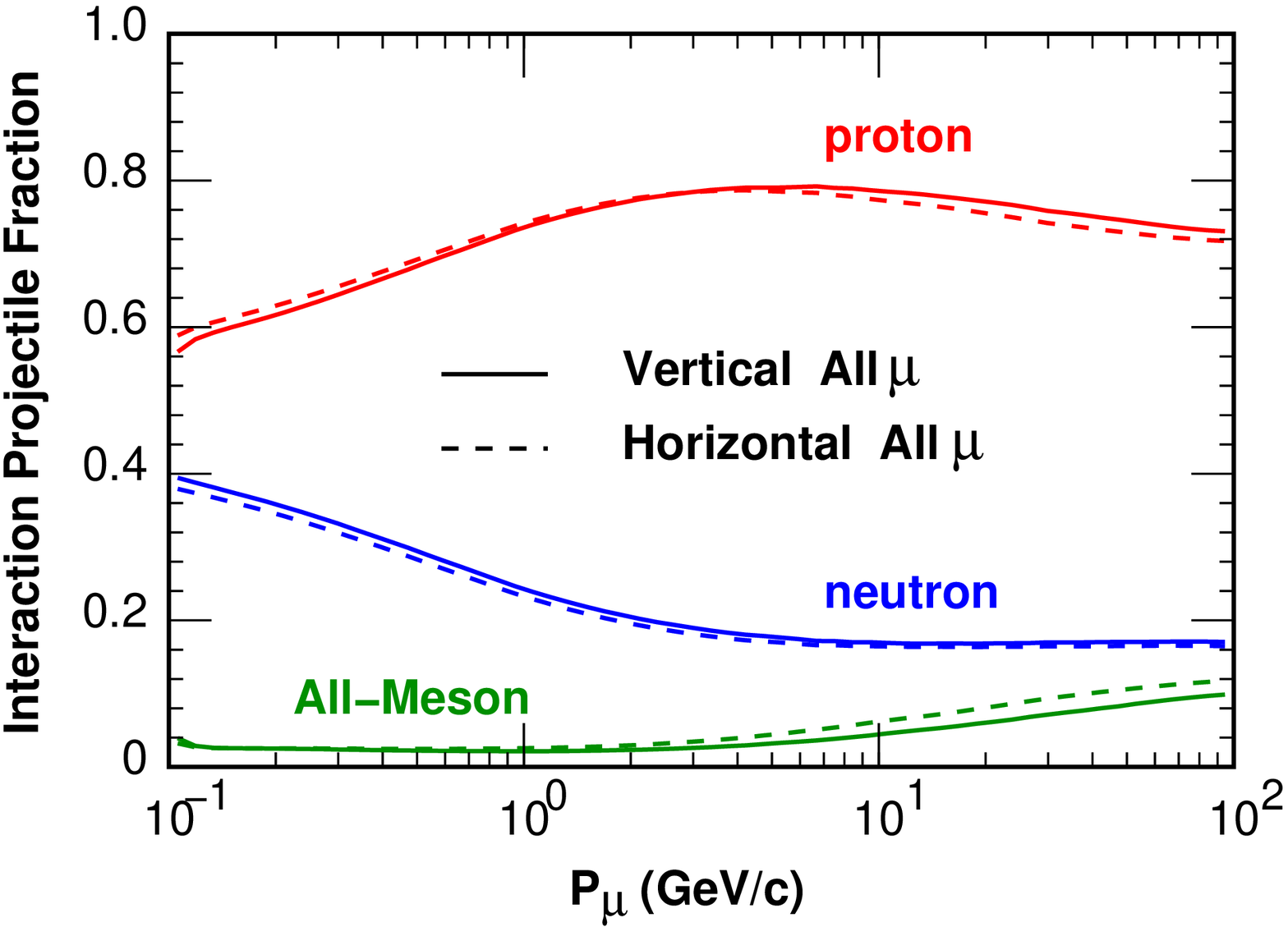}%
\caption{\label{projectile-lepton}
  Left panel; relative ratio of projectile particle ratio for vertical downward
  and horizontally moving atmospheric neutrino
  calculated for Kamioka summing all kind of neutrinos.
  Right panel; relative ratio of projectile particle ratio for vertical
  downward moving atmospheric muon calculated for Kamioka, summing both signed
  muons.
  }
\end{figure}

In Sec.~\ref{varint}, we assumed that the projectile flux
$\Phi_{proj}(N^{proj},p_N^{proj},x^{int})$
is not largely affected by the variation of the hadronic interactions.
We would like to comment on this assumption and the uncertainty due to it.
Classifying the projectile particles of the hadronic interaction with air nuclei
into three types, proton, neutron, and all as mesons as in Sec.~\ref{formulation},
we plot  the fraction of them for when their hadronic interaction resulted
in the target lepton production in Fig.~\ref{projectile-lepton},
summing all kind of neutrinos in the left panel,
and summing both signed muons in the right panel.

 The primary cosmic ray energy which produce the
 atmospheric neutrino and muon fluxes we are studying
 is less than a few TeV, and the  proton neutron ratio ($N_p/N_n$)
 is around 5.
 However, from Fig,.~\ref{projectile-lepton}, we find the
 $N_p/N_n$ ratio of the projectile particle  directly related to
 those atmospheric neutrino and muon fluxes is 1.5 at the lowest
 energy and around 4 in the highest energy  of our study in this paper.
 This means that  some of the projectile
particle have experienced hadronic interaction before
they create the parent meson
of the atmospheric neutrino and muon.
The  small $N_p/N_n$ ratio at low energies
means they are created by the projectiles which suffered  more
from the hadronic interaction than the atmospheric
neutrino and muon at higher energies.

As the contribution of mesons projectile is small for atmospheric neutrino and muon
below 100~GeV,  
we consider here the variation of  $N_p/N_n$ ratio only.
We repeat the study in section~\ref{variflx},
changing the $N_p/N_n$ ratio by $\pm 10$~\% for the atmospheric muon 
but fixing it to the original value for the atmospheric neutrino,
and fixing the $N_p/N_n$ ratio to the original value for the atmospheric muon
but changing it by $\pm 10$~\% for the atmospheric  neutrino by hand.
With those changes, we find the peak position of $\Delta \Phi_\nu/\Phi_\nu$
distribution moves only $\pm 2\sim 3$~\%,
which is would not be seen clearly if we add the distribution in Fig.~\ref{varinu}.
Note, the same change of $N_p/N_n$ ratio for the atmospheric muon 
and atmospheric neutrino shift the peak position to the opposite direction.
Considering the fact that a variation of hadronic interaction model
would change the  $N_p/N_n$ ratios for the 
atmospheric neutrino and muon fluxes to the same direction,
we may conclude that the possible variation of
the projectile particle ratio with the variation of hadronic interaction
model does not affect the former section analysis largely.

\begin{figure}[ht]
  \includegraphics[height=5.5cm]{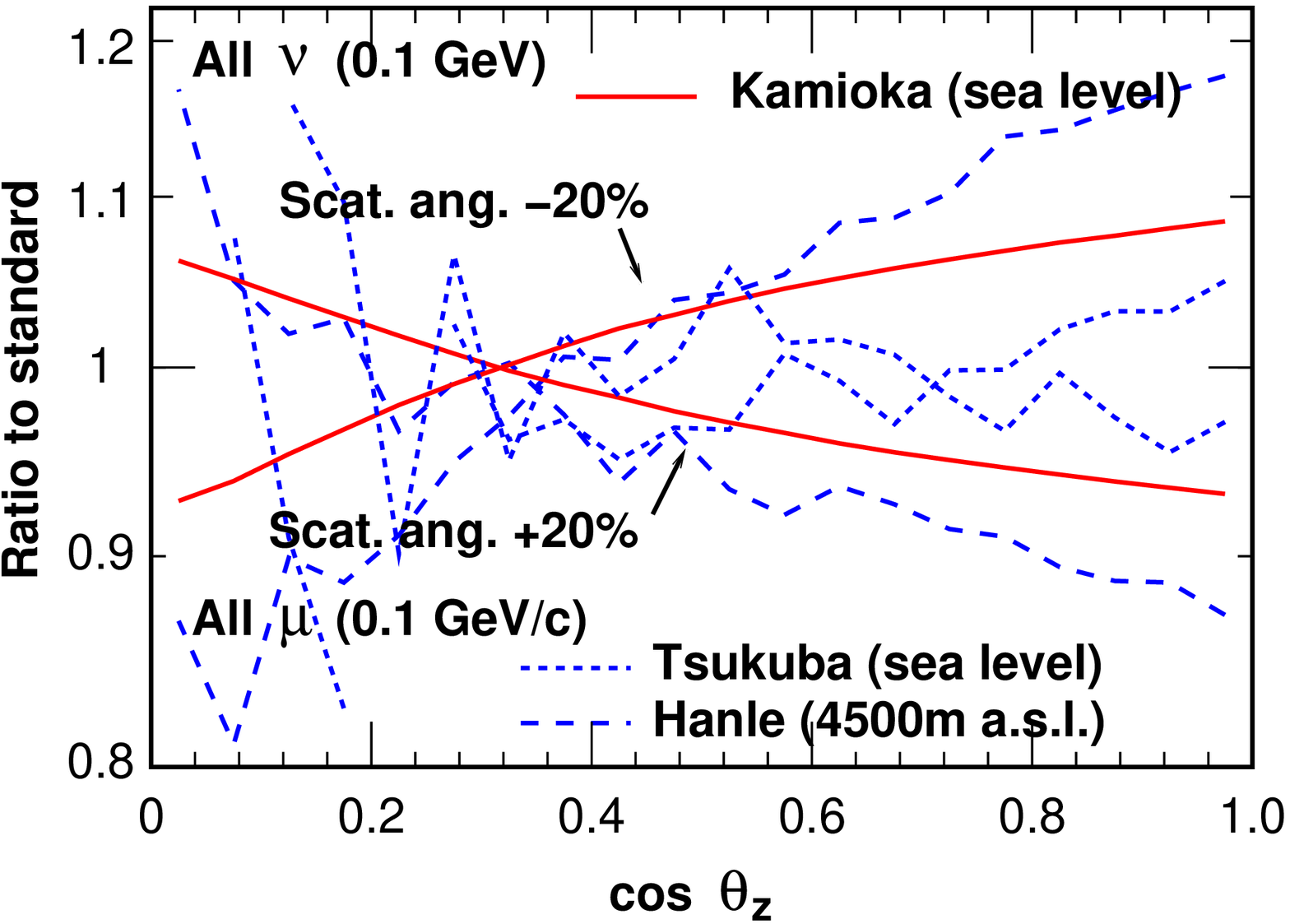}
  \includegraphics[height=5.5cm]{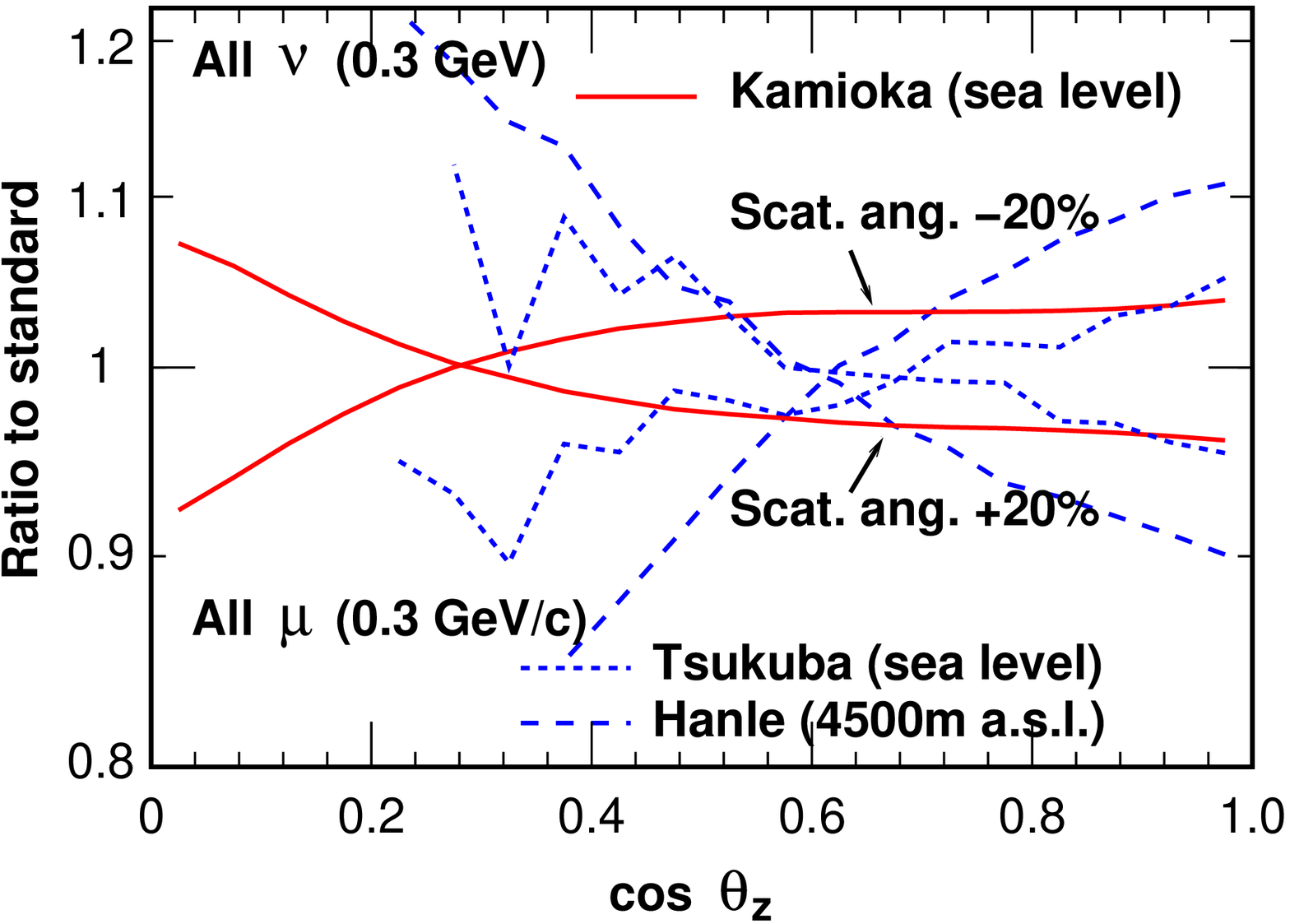}
\caption{\label{s-angle}
  The flux ratios of atmospheric neutrino and muon calculated with 20\% larger
  and 20\% smaller scattering angle to that with standard scattering angel
  in our calculation scheme at several sites.
  The solid lines are for the ratios of atmospheric neutrino flux calculated at Kamioka,
  dashed lines are the ratio of atmospheric muon flux calculated at Hanle,
  and dotted lines are the ratio of atmospheric muon flux calculated at Tsukuba.
  Left panel is for the atmospheric neutrino at $E_\nu =0.1$~GeV,
  and atmospheric muon with momentum at $P_\mu=0.1$~GeV$/$c.
  Right panel  is for the atmospheric neutrino at $E_\nu =0.1$~GeV
  and atmospheric muon at $P_\mu=0.1$~GeV$/$c.
}
\end{figure}

Another potentially  important source of the uncertainty for the
 lower energy atmosphere neutrino flux 
is the error of the scattering angle in the hadronic interaction.
It is well known that the three-dimensional calculation of atmospheric neutrino flux
shows an enhancement of the flux for near horizontal directions\cite{battis2002},
and is sometimes called the ``horizontal enhancement''.
This is due to the bending of secondary particles from the projectile particle
in the hadronic interaction, and is not seen
in the one-dimensional calculations.
Therefore, the error in the measurement of the scattering angle could result in the
error of the prediction of the low energy atmospheric neutrino flux.

To quantify the uncertainty of the atmospheric neutrino flux
due to the error in the scattering angle in the hadronic interaction,
we calculate the atmospheric neutrino flux at Kamioka and muon flux
at several observation sites changing the scattering angles of the
hadronic interaction model to
 $\pm$20~\% larger or smaller ones by hand.
Summing all kind of neutrino flux calculated at Kamioka,
we plot the ratio of the scattering angle changed neutrino fluxes to
the original one as the function of zenith angle
at $E_\nu= 0.1$~GeV (left panel) and at  $E_\nu= 0.3$~GeV (right panel)
in Fig.~\ref{s-angle}.
We also sum the flux of both signed atmospheric muon 
calculated at Tsukuba and Hanle,
and plot the ratio of the scattering angle changed muon fluxes to the
original one as the function of zenith angle 
at $P_\mu=0.1$~GeV$/$c (left panel) and at  $P_\mu= 0.3$~GeV$/$c (right panel)
in Fig.~\ref{s-angle}.
We find  $\sim \pm$ 10~\% variation for the scattering angle changed
neutrino flux at the near horizontal direction in both energy,
and for vertically downward direction at 0.1~GeV.
Therefore, if we want to reduce the error in the calculation
of atmospheric neutrino to $\sim$ 5~\%,
we need to reduce the uncertainty of the hadronic interaction
scattering angle to $\lesssim$ 10~\%.  
At the same time, 
We also observe that the same change of the scattering angle also cause a large the
change in the zenith angle dependence of atmospheric muon flux observed at 
high altitude site as Hanle (4500m a.s.l.),
but a smaller change at sea level (Tsukuba).
The muon observed at higher altitude as Hanle, whose the production altitude
is close to the observation altitude, are a little suffered from the
muon energy loss, and the change of scattering angle appears as a large
effect on the zenith angle dependence of  muon flux.
On the other hand, the muon observed at the sea level suffer
the maximum energy loss, and the observed zenith angle dependence is
that of higher energy one, where the change of the 
scattering angle appears smaller effect on the zenith angle
dependence of  muon flux.

The precision measurement of the scattering angle in the 
hadronic interaction is the work of accelerator experiments.
However, the Fig.~\ref{s-angle} shows a possibility to study
the uncertainty of it 
by measuring the atmospheric muon flux as the function of
the zenith angle.
We note that this observation must be carried out at high
mountain like Hanle (4500m a.s.l.).
As the atmospheric muon flux decreases quickly with the
zenith angle, the larger flux is preferable for this
observation.
We expect $\sim$~4 times larger atmospheric muon flux at Hanle
than that at sea level (see Fig.~\ref{mudata}),
Also the effect of the variation of the scattering angle
is more visible in the atmospheric muon flux data
observed at higher mountain.
We may reduce the uncertainty of
the scattering angle in hadronic interaction,
by reconstructing the atmospheric muon flux accurately
observed at Hanle.

 \section{Summary and Conclusions}

 Based on the pseudo-analytic formulation for the calculation
 of the atmospheric lepton flux, we developed a method to
 construct the variation of hadronic interaction model with the random
 numbers.
 Then
 we construct a huge number of the variation of the interaction model and
 the variation of atmospheric neutrino and 
 muon flux with them. 
 We find that  when we select the variation of interaction models whose
 calculated atmospheric muon fluxes is close to the original one,
 the atmospheric neutrino flux calculated with that is also close to the original one.
 By considering our interaction model, with which we are calculating
 the atmospheric neutrino and muon fluxes, is a variation of the
 ideal interaction model which can predict the true atmospheric
 neutrino and muon fluxes,
 we may conclude that  when we can  reconstruct the 
 atmospheric muon flux measured by a precision experiment,
 we can also calculate the atmospheric neutrino flux
 accurately.
 
 Note,  in our former studies, we modify the hadronic interaction
 model and reconstruct the accurately measured atmospheric muon flux
 in a good accuracy.
 However,  the study of this paper shows that there remains some uncertainty
 of the atmospheric neutrino flux depending on the observation site 
 and the minimum momentum of the atmospheric muon flux data
 rather than on the residual of the reconstruction.
 It is important to improve the muon observation equipment
 and find a better observation site for atmospheric muon flux.
 We hope the technology used in the recent primary cosmic ray
 observation detectors would improve also the muon observation
 detectors.
 For the observation site, we find that the atmospheric muon flux
 data observed at high  mountain is better than that observed at 
 a lower altitude site, to reduce the uncertainty of the
 atmospheric neutrino flux.
 It seems the  mountain site (3000 $\sim$ 5000 m a.s.l.) works most
 efficiently for this work, because the remaining uncertainty decrease
 with the altitude of the observation site up to 4500 m a.s.l.,
 but if we go up to the  balloon altitude  ($\sim$ 32 km),
 the remaining uncertainty rather  increases.

 As other source of the uncertainty of the atmospheric neutrino
 calculation, we considered the uncertainty of the projectile
 particle flux for the hadronic interaction which create the parent
 meson of the atmospheric neutrino and muon.
 We studied it by changing the relative ratio of the kind of projectile
 particles in the above variation study of the interaction model.
 However, the result is virtually the same.
 This is because the variation study of the hadronic interaction model
 cover the variation of the projectile particles flux.

 We observed that the uncertainty of the scattering angle in hadronic
 interaction is also the source of  uncertainty of the low energy
 atmospheric neutrino flux prediction due to the horizontal enhancement.
 This could be crucial to the study of neutrino physics,
 since this uncertainty result in the uncertainty in the
 zenith angular distribution of atmospheric neutrino flux.
 To study this uncertainty, we calculated the atmospheric
 neutrino flux, assuming the variation of the scattering angle by $\pm 20$~\%,
 and 
 we find the flux difference is $\lesssim$ 10~\% at 0.1 GeV and 0.3 GeV,
 both for the vertical downward and horizontal direction.
 If we reduce the uncertainty of the scattering angle  in the hadronic interaction
 to $\lesssim$ 10~\%, the  uncertainty of atmospheric neutrino would
 be $\lesssim$ 5~\%.
 The uncertainty of the scattering  should be studied at the accelerator experiment,
 but the  study of atmospheric muon zenith angle variation at high mountain altitude
 as Hanle, the atmospheric muon observation can also contribute to reduce it.

 Lastly, we would like to comment on the relation of our work and
 accelerator experiment in the calculation of  the atmospheric neutrino flux.
 First of all, we must confess that the interaction model we are using is
 basically constructed using the accelerator data.
 Without the acceleration experiments,
 we could not start the calculation of the atmospheric neutrino flux.
 We would like to note that the  accelerator experiment can improve
 the study of this paper, the reduction of the uncertainty of atmospheric
 neutrino flux using the accurately measured muon flux,
 We have assumed 50 \% uncertainty in the integral kernel density at each
 grid point for all kind of hadronic interaction related to the atmospheric
 neutrino and muon production.
  If we can start with much smaller uncertainty for the integral kernels,
  the remaining uncertainty would be smaller.
  Although it is a higher energy problem, the kaon production uncertainty is
  typically this case.
  We believe the cooperation with accelerator study is necessary to achieve 
  much higher accuracy in the prediction of the atmospheric neutrino flux.

 \begin{acknowledgments}
We are greatly appreciative to Jun Nishimura 
for the helpful discussions and comments through this paper.
We are grateful to Sadakazu Haino and Anatoli Fedynitch 
for the discussion.
We also thank the ICRR of the University of Tokyo, 
especially for the use of the computer system.
This work was carried out under the support of Kakenhi.
 \end{acknowledgments}
 \bibliography{mnflux-2018}
\end{document}